%% file: article.tex
\documentclass[letterpaper, 11pt]{article}

\usepackage[brazilian,hyperpageref]{backref}
\usepackage[margin=0.8in]{geometry}
\usepackage[utf8]{inputenc}	
\usepackage[table]{xcolor}
\usepackage[T1]{fontenc}
\usepackage{amsthm,amsmath,vmargin,url}
\usepackage{graphicx,booktabs}
\usepackage{indentfirst}	
\usepackage{longtable}
\usepackage{enumerate}
\usepackage{hyperref}
\usepackage{verbatim}
\usepackage{amsfonts}
\usepackage{makecell}
\usepackage{multirow}
\usepackage{lipsum}
\usepackage{helvet}
\usepackage{subfig}
\usepackage{natbib}
\usepackage{tikz}
\usepackage{floatrow}
\DeclareMathOperator*{\tr}{tr}
\DeclareMathOperator*{\diag}{diag}

\DeclareMathOperator*{\GCV}{GCV}

\newcommand{\normal}{\mathrm{N}}
\newcommand{\prob}{\mathrm{P}}
\newcommand{\unif}{\mathrm{U}}
\newcommand{\ber}{\mathrm{Ber}}

\newcommand{\dBeta}{\mathrm{Beta}}
\newcommand{\GI}{\mathrm{IG}}
\newcommand{\E}{\mathrm{E}}

\renewcommand{\vec}[1]{\boldsymbol{#1}}
\usetikzlibrary{arrows}
\usetikzlibrary{matrix}
\usetikzlibrary{positioning}

\begin{document}
\title{\textbf{Bayesian Adaptive Selection of Basis Functions for Functional Data Representation}}
\author{Pedro Henrique T. O. Sousa$^{1*}$, Camila P. E. de Souza$^{2}$, Ronaldo Dias$^{1}$}
\date{}
\maketitle

\begin{small}
\noindent $^{1}$Department of Statistics, University of Campinas, SP, Brazil \\
\noindent $^{2}$Department of Statistical and Actuarial Sciences, University of Western Ontario, ON, Canada\\
\vspace{0.2cm}
\noindent *Correspondence: phtos\_est@gmail.com
\end{small}

\begin{abstract}
				Considering the context of functional data analysis, we developed and applied a new Bayesian approach via Gibbs sampler to select basis functions for a finite representation of functional data. The proposed methodology uses Bernoulli latent variables to assign zero to some of the basis function coefficients with a positive probability. This procedure allows for an adaptive basis selection since it can determine the number of bases and which should be selected to represent functional data. Moreover, the proposed procedure measures the uncertainty of the selection process and can be applied to multiple curves simultaneously. The methodology developed can deal with observed curves that may differ due to experimental error and random individual differences between subjects, which one can observe in a real dataset application involving daily numbers of COVID-19 cases in Brazil. Simulation studies show the main properties of the proposed method, such as its accuracy in estimating the coefficients and the strength of the procedure to find the true set of basis functions. Despite having been developed in the context of functional data analysis, we also compared the proposed model via simulation with the well-established LASSO and Bayesian LASSO, which are methods developed for non-functional data.

\end{abstract} 

\input{introduction}

\input{method}

\input{simulations}

\input{real_data}

\input{discussion}

\section*{Acknowledgments}

This research work is supported by the Coordenação de Aperfeiçoamento de Pessoal de Nível Superior (CAPES) and the Natural Sciences and Engineering Research Council of Canada (NSERC).

\bibliographystyle{asa}
\nocite{cezar,reviewFDA,mcs,fdaReMatlab,infFDA,cuevas,Ferraty,Ferraty2,Handwriting,hmafd}
\bibliography{main}

\newpage
\input{appendix}

\end{document}

%% file: introduction.tex
\section{Introduction}
Although the history of the functional data field goes back to the 1950s with \citeauthor{Grenander} (\citeyear{Grenander}) and \citeauthor{Rao} (\citeyear{Rao}), the term functional data analysis (FDA) itself was coined in \citeyear{ADF} by \citeauthor{ADF}, and has since gained the attention of researchers from different science fields due to its particular and innovative characteristics. The term “functional” emphasizes that statistical methods in this area take into account the functional nature of the data.

With the progress of computational tools and other technologies, data can be more easily measured in a vast and dense grid of points from a continuous domain, making it common to analyze and interpret some data as curves \citep{mcs,ReS,Ferraty2}. FDA applications can be found in studies involving economic data (\citeauthor{ecoFDA}, \citeyear{ecoFDA}), imaging data (\citeauthor{Locantore}, \citeyear{Locantore}; \citeauthor{enescfd}, \citeyear{enescfd}), medical data \citep{medicalFDA,shi2021functional}, data on electricity consumption \citep{dias,Lenzi,Camila,franco2021aggregated},  biologging data \citep{fu2019model,sidrow2021modelling}, and spectral data \citep{dias2015aggregated, robbiano2016improving}, among others.  


The idea of functional data analysis is based on the assumption that data are single entities rather than mere sequences of individual observations (\citeauthor{ReS}, \citeyear{ReS}). In other words, functional data are characterized by a collection of functions and as such are intrinsically infinite dimensional. As a consequence of the infinite-dimensional structure, complete and continuous disclosure of functional data to researchers through observation becomes impossible. In practice, the data are not observed at every evaluation point $t$, as this would lead to an uncountable number of values, but it is densely recorded at discrete points. Thus, functional data are usually observed and recorded in $n$ pairs of the type ($t_{j}$,$y_{j}$), in which $y_{j}$ is an observed value of the functional $x$ at the point $t_{j}$ and it is possibly disturbed by noise. The evaluation  points, $t_{j}$s, can represent time, or any other continuous variable of interest \citep{ReS}. Therefore, we assume there is a functional $x$ that is smooth and differentiable. This functional is not observed in its entirety, but gives rise to observed data so that:
\begin{equation}
y_{j}=x(t_{j})+\epsilon_{j}\text{,}
	\label{modelosimples}
\end{equation} where the $\epsilon_{j}$s represent the random noises that contaminate the observations and mask the smooth trajectory of the functional. We can rewrite Expression \eqref{modelosimples} in vector notation as
\begin{equation}
\vec{y}=x(\vec{t})+\vec{\epsilon}\text{,}
\label{modelosimplesvec}
\end{equation}where  $\vec{y}=(y_{1},\;y_{2},\dots,y_{n})^{'}$ and $\vec{\epsilon}=(\epsilon_{1},\epsilon_{2},\dots,\epsilon_{n})^{'}$ and $x(\vec{t})$ is the vector with the values of function $x$ evaluated at the points $t_{1},t_{2}, \ldots , t_{n}$.


Since the observed data are represented by the Expression \eqref{modelosimplesvec}, the conversion of data (finite) into functions (which theoretically can be evaluated in an infinite number of points) will necessarily involve some type of smoothing, also called regularization (\citeauthor{fdaReMatlab}, \citeyear{fdaReMatlab}). 
The need for regularization is to avoid overfitting, meaning to prevent the functional curve from being too sinuous and biased towards interpolation. 

A common strategy in FDA to deal with high dimensionality and the consequent need for smoothing is to use basis function expansion to represent functions. Basis functions are the building blocks of functional data analysis and typically determine the mechanism by which regularization is enforced. As such an expansion is characterized by the linear combination of known basis functions as follows:
\begin{equation}
	x(t)=\sum_{k=1}^{K}\beta_{k}B_{k}(t)=\vec{\beta}^{'}\vec{B}(t)\text{,}\nonumber
\end{equation}

\noindent where $\vec{\beta}=(\beta_{1},\;\beta_{2},\dots,\;\beta_{K})^{'}$ is the vector of unknown coefficients of the combination and $\vec{B}(t)=(B_{1}(t),\;B_{2}(t),\dots,\;B_{K}(t))^{'}$ is the vector of basis function evaluations. Thus, smoothing is determined by the number of basis functions $K$. The smaller $K$ is, the greater is the regularization. Estimates for the expansion coefficients are usually obtained by the Ordinary Least Squares (OLS) method (\citeauthor{ReS}, \citeyear{ReS}).

In some cases, the regularization described above does not present good performance because of the need to use many basis functions. Therefore, another approach for these cases is the smoothing splines method (\citeauthor{wahba1990spline}, \citeyear{wahba1990spline}; \citeauthor{nrGLM}, \citeyear{nrGLM}; \citeauthor{tesl}, \citeyear{tesl}), which penalizes the curvature of the function. There are other tools for data smoothing in addition to smoothing splines and OLS estimation via basis expansion, such as kernel smoothing and the local least squares technique (\citeauthor{KS}, \citeyear{KS}; \citeauthor{LPMA}, \citeyear{LPMA}; \citeauthor{nrss}, \citeyear{nrss}; \citeauthor{pgs}, \citeyear{pgs}). 

Whether the estimation is by OLS via basis expansion or the aforementioned smoothing methods, and regardless of the capacity that each has to provide good fitting while regularizing, none of them can perform basis selection. In short, there is a demand for more flexible methods that perform basis selection and even automatically adapt to a larger collection of curves.


In the statistical literature on regularization of functional data, it is noted that successful methods have been developed to build new systems of basis functions or adaptive splines that fit different curve shapes \citep{pgs}. However, not every one focuses on developing or adapting models that are capable of providing greater flexibility with existing foundations.

A different but related problem is variable selection in linear models. There is a wide range of  selection methods available for these models, including RIDGE (\citeauthor{ridge}, \citeyear{ridge}), LASSO (\citeauthor{lasso}, \citeyear{lasso}), and its variants such as elastic net (\citeauthor{elasticnet}, \citeyear{elasticnet}), group LASSO (\citeauthor{groupLASSO}, \citeyear{groupLASSO}), sparse group LASSO (\citeauthor{sparsegroupLASSO}, \citeyear{sparsegroupLASSO}), and overlap group LASSO (\citeauthor{overlapGRlasso}, \citeyear{overlapGRlasso}), among others. There are also several Bayesian methods for covariate selection. One of the pioneering Bayesian methods for covariate selection, known as SSVS (Stochastic Search Variable Selection), was introduced by \citeauthor{McCulloch93} (\citeyear{McCulloch93}). In this method, the prior distribution for the coefficient associated with the covariate is a mixture of two normal distributions with different variances. Later, the creators of SSVS extended the method to the field of generalized linear models (\citeauthor{McCulloch96}, \citeyear{McCulloch96}), just before \citeauthor{Brown} (\citeyear{Brown}) adapted SSVS for the multivariate case.

Despite introducing the idea of variable selection, the SSVS is not able to zero out the coefficients of the linear model, so the selection of variables is given through an evaluation of the posterior joint distribution so that it is possible to identify which coefficients are significantly different from zero. To overcome this problem, \citeauthor{Kuo} (\citeyear{Kuo}) presented a Bayesian model characterized by incorporating $2^p$ submodels through introducing an auxiliary indicator variable that is independent of the coefficient vector and makes it possible to assign zero to certain model coefficients with a positive probability. Some alternative Bayesian models inspired by the previously mentioned models were also proposed, as can be seen in \citeauthor{Carlin} (\citeyear{Carlin}) and \citeauthor{Dellaportas} (\citeyear{Dellaportas}).

Finally, a Bayesian method for variable selection that deserves to be highlighted is the one proposed by \citeauthor{art:bl-Casella}, (\citeyear{art:bl-Casella}). Their method assigns a Laplace distribution to the coefficients which not only deals with the selection of variables, but is also equivalent to the LASSO frequentist method when the mode is used as a summary measure of the posterior values. Although the method establishes a link with LASSO through the mode as a summary measure, the same does not happen if the median or the mean are used to summarize the posterior information, in such a way that Bayesian LASSO (as the method is known) is not able to assign zero directly to coefficients for these cases.

Among all the frequentist and Bayesian models developed for applications in linear models, the model proposed by \citeauthor{Kuo} (\citeyear{Kuo}) presents interesting attributes for adaptation and consequent application in data with functional structure. From a frequentist perspective, \cite{cezar} considered latent auxiliary indicator variables as in \cite{Kuo} and performed basis selection via a modified stochastic Expectation-Maximization (SEM) algorithm \citep{celeux1992stochastic}. 

In this work, we propose a Bayesian model aimed at functional data application, which is based on the concept introduced by \citeauthor{Kuo} (\citeyear{Kuo}) of considering a Bernoulli random variable for each coefficient of the linear basis expansion. Our proposed model can promote the selection of basis functions for multiple curves simultaneously. Our strategy consists of formulating a hierarchical model to provide the necessary full conditional distributions for implementing the Gibbs sampler, which will generate posterior samples for estimating summary statistics. Our method is implemented in R and is available at \url{https://github.com/phtosEST/Adaptive-Bayesian-Selection-of-Basis-Functions}.



Section \ref{cap2} of this work describes our proposed Bayesian methodology for basis function selection for functional data representation. Section \ref{cap3} presents the design and consequent results of the various simulation studies carried out to assess the performance of the proposed approach. A brief description of the comparison made with LASSO and Bayesian LASSO is also presented. More details on this comparison can be found in the Supplementary Material. Section \ref{realdatacap} presents the application results of our proposed model for selecting bases and the consequent curve fitting of new cases of COVID-19 in each state in Brazil, including the Federal District.

%% file: method.tex
\section{Methods}
\label{cap2}

For the introduction of our proposed Bayesian methodology, suppose that there are $m$ curves with $n_{i}$ observations at the points $t_{ij}\in A\subseteq \mathbb{R}$, in which $i \in \{1,2,\dots,m\}$ and $j \in \{1,2,\dots,n_{i}\}$. So, consider that: 
\begin{equation}
	y_{ij}=g(t_{ij})+\epsilon_{ij}\text{,}
	\label{funG}
\end{equation}
\noindent where $\epsilon_{ij}$ is a random component with a normal distribution with zero mean and variance equal to $\sigma^2$.
While $y_{ij}$ represents the observed value of the $i$th curve at the point $t_{ij}$, the term $g(t_{ij})$ can be expressed as a linear random combination of basis functions so that: 
\begin{equation}
	g(t_{ij})=\sum_{k=1}^{K}Z_{ki}\beta_{ki}B_{k}(t_{ij})\text{,}
	\label{eq:g}
\end{equation} where the $Z_{ki}$s are the  Bernoulli independent latent random variables with $\prob(Z_{ki}=1)=\theta_{ki}$ inducing basis selection. For a given fixed $i$, the vector with the final coefficients from the linear combination of bases is given by $\vec{\nu_{i}}=(\nu_{1i},\;\nu_{2i},\dots,\;\nu_{Ki})^{'}=(Z_{1i}\beta_{1i},\;Z_{2i}\beta_{2i},\dots,\;Z_{Ki}\beta_{ Ki})^{'}$. 
For the sake of simplicity the components $\nu_{ki}$s will be denoted coefficients, while the $\beta_{ki}$s will be called partial coefficients. 

Let 
\begin{gather}
	\vec{Y}=(y_{11},\;y_{12},\dots,\;y_{1n_{1}},\;y_{21},\;y_{22},\dots,\;y_{2n_{2}},\dots,\;y_{m1},\;y_{m2},\dots,\;y_{mn_{m}})^{'}\text{,} \nonumber \\
	\vec{\theta}=(\theta_{11},\;\theta_{21},\dots,\;\theta_{K1},\;\theta_{12},\;\theta_{22},\dots,\;\theta_{K2},\dots,\;\theta_{1m},\;\theta_{2m},\dots,\;\theta_{Km})^{'}\text{,}
	\nonumber \\
	\vec{\beta}=(\beta_{11},\;\beta_{21},\dots,\;\beta_{K1},\;\beta_{12},\;\beta_{22},\dots,\;\beta_{K2},\dots,\;\beta_{1m},\;\beta_{2m},\dots,\;\beta_{Km})^{'}\text{ and } \nonumber 
	\\
	\vec{Z}=(Z_{11},\;Z_{21},\dots,\;Z_{K1},\;Z_{12},\;Z_{22},\dots,\;Z_{K2},\dots,\;Z_{1m},\;Z_{2m},\dots,\;Z_{Km})^{'}\text{.} \nonumber
\end{gather}

\noindent Considering (\ref{funG}) and (\ref{eq:g}), we can build the following Bayesian hierarchical model:
\begin{gather}
	y_{ij}|\vec{Z},\vec{\beta},\sigma^2\sim\normal\left(\sum_{k=1}^{K}Z_{ki}\beta_{ki}B_{k}(t_{ij}),\sigma^{2}\right)\text{;}\nonumber\\
	\beta_{ki}|\sigma^{2},\tau^{2}\sim\normal(0,\sigma^2\tau^{2})\text{;}\nonumber\\
	Z_{ki}|\vec{\theta}\sim\ber(\theta_{ki})\text{;}\nonumber\\
	\theta_{ki}\sim\dBeta\left(\mu_{ki},1-\mu_{ki}\right)\text{;}\nonumber\\
	\tau^{2}\sim\GI(\lambda_{1},\lambda_{2})\quad \text{and}\quad\sigma^{2}\sim\GI(\delta_{1},\delta_{2})\text{.}
	\label{modeloRaiz}
\end{gather}

\noindent From this model and taking into account the following posterior joint density: 
\begin{equation}
		f(\vec{\beta},\vec{Z},\vec{\theta},\sigma^2,\tau^{2}|\vec{Y})\propto\pi(\sigma^2)\pi(\tau^2)\pi(\vec{\theta})f(\vec{\beta}|\tau^2,\sigma^2)p(\vec{Z}|\vec{\theta})f(\vec{Y}|\vec{\beta},\vec{Z},\sigma^2)\text{;}\nonumber
\end{equation}
\noindent it is possible to compute all full conditional distributions for the construction of the Gibbs sampler (\citeauthor{expGibbs}, \citeyear{expGibbs}).

In the model described in (\ref{modeloRaiz}), $\mu_{ki}$ is the hyperparameter of the $\theta_{ki}$ density, being also the mean of the respective prior distribution. As an initial assumption of the model, the $\beta_{ki}$ prior has a variance equal to $\sigma^2\tau^{2}$. While $\tau^{2}$ performs as a regularizing component, the inclusion of $\sigma^2$ establishes a dependency between $\beta_{ki}$ and $\sigma^2$ that is intuitive, desired and widely used in the statistical literature (\citeauthor{Hoff}, \citeyear{Hoff}; \citeauthor{art:bl-Casella}, \citeyear{art:bl-Casella}). Furthermore, the inclusion of $\sigma^2$ in this prior ensures that $\sigma^2$ does not appear in the full conditional averages of the $\beta_{ki}$'s. 

 Bayesian variable selection methods generally set priors for the coefficients as mixtures of two components: a spike concentrated around zero and a flat slab. For the spike, there are two most common specifications: an absolutely continuous spike distribution and a spike characterized by a point mass at zero (\citeauthor{OHara}, \citeyear{OHara}; \citeauthor{Gertraud}, \citeyear{Gertraud}).

When there is a mixture using an absolutely continuous spike distribution, a separate procedure is necessary to assess whether the estimate of a given coefficient is far enough from zero so that the respective coefficient is selected or not. \citeauthor{McCulloch93} (\citeyear{McCulloch93}) presented a method in which the coefficients of the priors are mixtures of this type. In this case, the mixtures' weights are defined by the latent random variables so that there is a statistical dependence between the coefficients and the latent variables.

In our proposed model, the prior distribution for each coefficient $\nu_{ki}=Z_{ki}\beta_{ki}$ from the linear combination of basis functions is a mixture of a point of mass distribution at zero with probability $1-\theta_{ki}$ and a normal $\normal(0,\sigma^{2}\tau^{2})$ with probability $\theta_{ki}$. In this case, the basis selection is made directly, since the estimate of a certain coefficient can be precisely equal to zero. Thus, there is no need for a subsequent procedure to the Gibbs sampler to complete the selection. Evidently, there is statistical independence between the latent random variables $Z_{ki}$s and the partial coefficients $\beta_{ki}$s, but the latent variables and the $\nu_{ki}$s (which are the true coefficients from the linear combination of basis functions) continue to be statistically dependent.

Considering the hierarchical model in \eqref{modeloRaiz}, it is still possible to add one more layer to the model by considering $\mu_{ki}$ as a parameter instead of a hyperparameter. In this case, the model can be defined as:
\begin{gather}
	y_{ij}|\vec{Z},\vec{\beta},\sigma^2\sim\normal\left(\sum_{k=1}^{K}Z_{ki}\beta_{ki}B_{k}(t_{ij}),\sigma^{2}\right)\text{;}\nonumber\\
	\beta_{ki}|\sigma^{2},\tau^{2}\sim\normal(0,\sigma^2\tau^{2})\text{;}\nonumber\\
	Z_{ki}|\vec{\theta}\sim\ber(\theta_{ki})\text{;}\nonumber\\
	\theta_{ki}|\mu_{ki}\sim\dBeta\left(\mu_{ki},1-\mu_{ki}\right)\text{;}\nonumber\\
	\mu_{ki}\sim \unif(0,\psi), \quad\text{em que}\quad \psi<1\text{;}\nonumber\\
	\tau^{2}\sim\GI(\lambda_{1},\lambda_{2})\quad \text{and}\quad\sigma^{2}\sim\GI(\delta_{1},\delta_{2})\text{.}
	\label{modeloAuto}
\end{gather}

Now, with $\mu_{ki}$ as a parameter, it is recommended to define $\psi$ with a value not so close to one. Since high values of $\mu_{ki}$  can pose numerical issues. The reason of this inconvenience is found in the posterior distribution of $\theta_{ki}$ since, given that $Z_{ki}=1$, such distribution converges to a degenerate beta when $\mu_{ki}$ tends to one. Furthermore, when considering $\mu_{ki}$ as a parameter, the posterior distribution of $\mu_{ki}$ is highly concentrated around one when $\theta_{ki}$ is close to one. These two behaviors combined can induce a vicious cycle in the Gibbs sampler, resulting in a model that although fits well, is not able to perform variable selection. Fortunately, this issue only arises when considering $\mu_{ki}$ as a parameter, since $\mu_{ki}$ is fixed for the model in (\ref{modeloRaiz}).

\subsection{Prior values for \texorpdfstring{$\theta_{ki}$}{}}

\begin{figure}[!htb]
	\centering
	\includegraphics[scale=0.6]{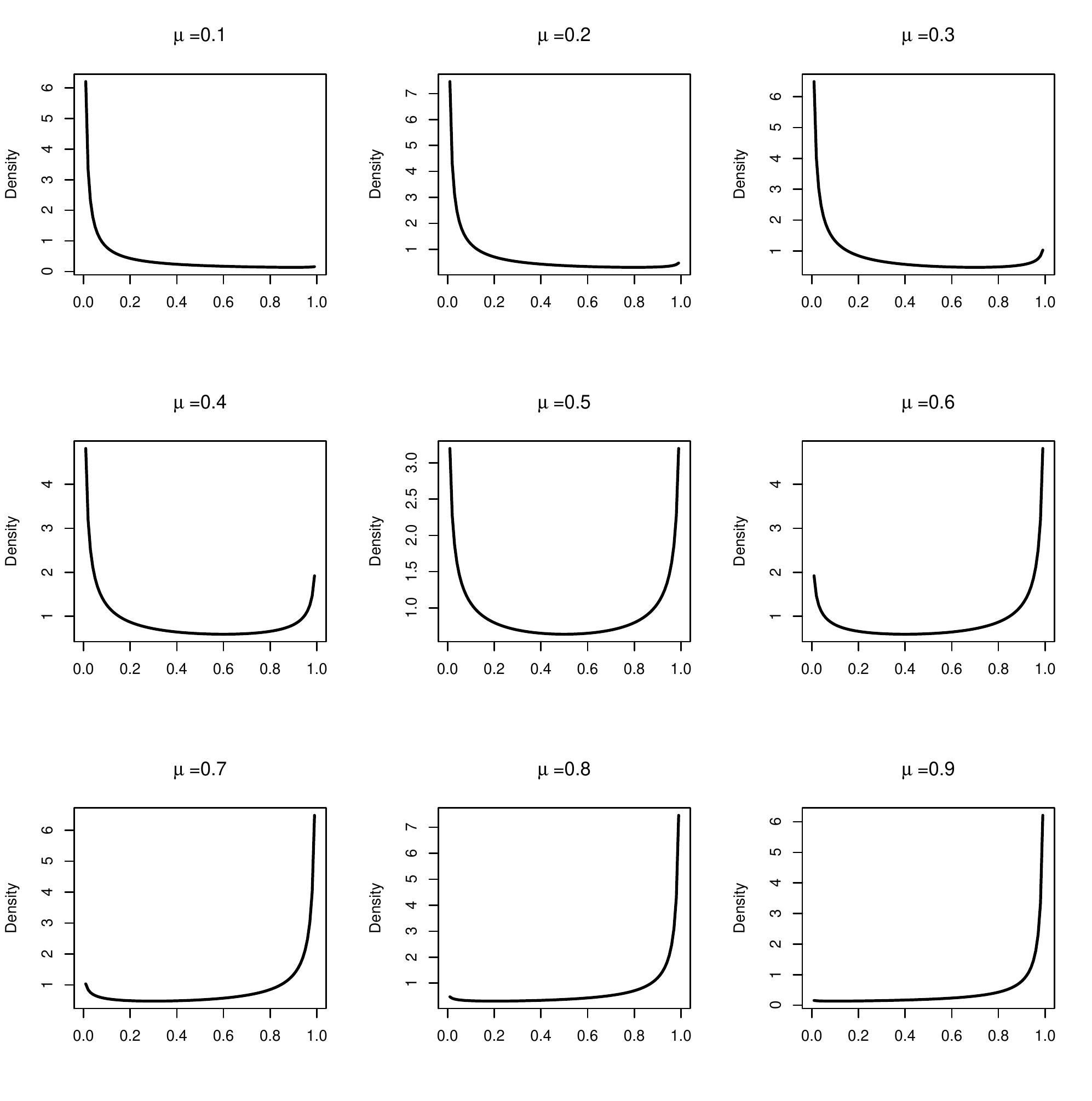}
	\caption{Some possible configurations for the $\theta_{ki}$ prior.}
	\label{fig:img2}
\end{figure}

For a brief evaluation of the $\theta_{ki}$ prior behavior as a function of $\mu_{ki}$, Figure \ref{fig:img2} presents some possible configurations for this density. The closer $\mu_{ki}$ is to $0.5$, the less informative the $\theta_{ki}$ prior is, since the respective density will be symmetric with a mean of 0.5 and a maximum variance.

Although the $\theta_{ki}$ prior explicitly presents only one hyperparameter, one can rewrite $\mu_{ki}$ so that for all $k$ and $i$, $\mu_{ki}=\frac{C}{K}$, where $C$ is a tuning parameter and can be interpreted as a prior guess about how many bases on average one would like to select. Through this perspective, the $\theta_{ki}$ prior has two hyperparameters, $C$ and $K$. To visualize this, just take $i$ fixed and consider the prior mean of the binomial generated by the sum of the $Z_{ki}$s: 
\begin{equation}
	\E\left(\sum_{k=1}^{K} Z_{ki}\right)=\sum_{k=1}^{K}\E\left(\E\left(Z_{ki}|\theta_{ki}\right)\right)=\sum_{k=1}^{K}\E(\theta_{ki})=\sum_{k=1}^{K}\mu_{ki}=\sum_{k=1}^{K}\frac{C}{K}=K\frac{C}{K}=C\text{.}\nonumber
\end{equation}

\subsection{Posterior joint density}
The hierarchical formulation thus constructed makes it possible to calculate the full conditional distributions of the model parameters, as well as the latent variables involved. Through the hierarchical formulation \eqref{modeloAuto}, the posterior joint density of $\sigma^2,\tau^2,\vec{\mu},\vec{\theta},\vec{\beta}$ and the latent variables $\vec{Z}$ is given by: 

\begin{gather}
f(\sigma^2,\tau^2,\vec{\mu},\vec{\theta},\vec{\beta},\vec{Z}|\vec{Y}) \propto	\pi(\sigma^2)\pi(\tau^2)\pi(\vec{\mu})f(\vec{\theta}|\vec{\mu})f(\vec{\beta}|\tau^2,\sigma^2)p(\vec{Z}|\vec{\theta})f(\vec{Y}|\vec{\beta},\vec{Z},\sigma^2)\nonumber\\
	\propto\left(\frac{1}{\sigma^2}\right)^{\delta_{1}+1}\exp\left\{-\frac{\delta_{2}}{\sigma^2}\right\}\left(\frac{1}{\tau^2}\right)^{\lambda_{1}+1}\exp\left\{-\frac{\lambda_{2}}{\tau^2}\right\}I_{(0,\psi)}(\mu_{ki})\frac{1}{\psi}\times\nonumber\\
	\left[\prod_{k=1}^{K}\prod_{i=1}^{m}
	(\theta_{ki})^{\mu_{ki}-1}(1-\theta_{ki})^{(1-\mu_{ki})-1}\left(\frac{1}{2\pi\sigma^2\tau^2}\right)^{\frac{1}{2}}\exp\left\{-\frac{\beta_{ki}^{2}}{2\sigma^2\tau^2}\right\}(\theta_{ki})^{Z_{ki}}(1-\theta_{ki})^{1-Z_{ki}}\right]\nonumber\\	
	\left(\frac{1}{2\pi\sigma^2}\right)^{\frac{\sum_{i=1}^{m}n_{i}}{2}}\exp\left\{-\frac{\sum_{i=1}^{m}\sum_{j=1}^{n_{i}}(y_{ij}-g(t_{ij}))^2}{2\sigma^2}\right\}\text{.}\nonumber
\end{gather}

The logarithm of this posterior joint density is then up to a constant, equal to: 
\begin{gather}
	-(\delta_{1}+1)\log(\sigma^2)-(\lambda_{1}+1)\log(\tau^2)-\frac{\delta_{2}}{\sigma^2}-\frac{\lambda_{1}}{\tau^2}-\left[I_{(0,\psi)}(\mu_{ki})\log(\psi)\right]+\nonumber\\
	\left[\sum_{k=1}^{K}\sum_{i=1}^{m}(\mu_{ki}+Z_{ki}-1)\log\left(\frac{\theta_{ki}}{1-\theta_{ki}}\right)\right]-\frac{mK}{2}\log\left(2\pi\sigma^2\tau^2\right)-\left(\frac{\sum_{k=1}^{K}\sum_{i=1}^{m}\beta_{ki}^{2}}{2\sigma^2\tau^2}\right)-\nonumber\\
	\frac{\sum_{i=1}^{m}n_{i}}{2}\log(2\pi\sigma^2)-\frac{\sum_{i=1}^{m}\sum_{j=1}^{n_{i}}(y_{ij}-g(t_{ij}))^2}{2\sigma^2}\text{.}\nonumber
\end{gather}

\subsection{Gibbs Sampler}

In Appendix \ref{A_full}, we derive in detail the full conditional distributions for the model given in \eqref{modeloAuto} used in our Gibbs sampler. The algorithm is analogous to the model in \eqref{modeloRaiz}. 

The panel below presents the standard procedure for implementing the Gibbs sampler in a clear and summarized way.\\[0.3cm]

\noindent\fbox{\begin{minipage}{15cm} 
\textcolor{white}{..}1. Define the hyperparameters;\\
\textcolor{white}{..}2. Assign initial chain values to parameters;\\
\textcolor{white}{..}3. For $\text{c}=2,\dots,\text{Nint}$:\\
\textcolor{white}{..}3.1 Sample ${\sigma^{2}}^{(c)}\sim f(\sigma^2|\vec{\beta}^{(c-1)},{\tau^2}^{(c-1)},\vec{Z}^{(c-1)},\vec{Y})$, an Inverse-Gamma as in \eqref{cmpos_sigma2}; \\
\textcolor{white}{..}3.2 Sample ${\tau^{2}}^{(c)}\sim f(\tau^2|\vec{\beta}^{(c-1)},{\sigma^2}^{(c)})$, an Inverse-Gamma as in \eqref{cmpos_tau2}; \\
\textcolor{white}{..}3.3  For $i=1,\dots,m$:\\
\textcolor{white}{..}3.3.1 For $k=1,\dots,K$:\\
\textcolor{white}{..}3.3.1.1 Sample $\mu_{ki}^{(c)}\sim f(\mu_{ki}|\theta_{ki}^{(c-1)})$, a continuous Bernoulli as in \eqref{cmpos_mu};\\
\textcolor{white}{..}3.3.1.2 Sample $Z_{ki}^{(c)}\text{ com }\prob(Z_{ki}=1|\vec{\beta}^{(c-1)},\theta_{ki}^{(c-1)},{\sigma^2}^{(c)},\vec{Z_{-[ki]}}^{(*)},\vec{Y})$ as in \eqref{cmpos_Z};\\
\textcolor{white}{..}3.3.1.3 Sample $\theta_{ki}^{(c)}\sim f(\theta_{ki}|\mu_{ki}^{(c)},Z_{ki}^{(c)})$, a beta as in \eqref{cmpos_theta};\\
\textcolor{white}{..}3.4  For $i=1,\dots,m$:\\
\textcolor{white}{..}3.4.1 Sample $\vec{\beta_{.i}}^{(c)}\sim f(\vec{\beta_{.i}}|{\sigma^2}^{(c)},{\tau^2}^{(c)},\vec{Z}^{(c)},\vec{Y})$, a multivariate normal as in \eqref{cmpos_beta}.\\
\end{minipage}}\\

The term $\vec{Z_{-[ki]}}^{(*)}$ represents a mixed vector with components from the previous iteration that have not yet been updated and components that have already been updated in the current iteration. As $i$ and $k$ change in 3.3 and 3.3.1, respectively, the vector $\vec{Z_{-[ki]}}^{(*)}$ is updated. 
Therefore, at the end of this process with the convergence of the chain, we obtain a sample of the posterior joint distribution.

%% file: simulations.tex
\section{Studies with synthetic data and simulations}
\label{cap3}

\subsection{Synthetic data studies and simulation design}
\label{cap3seção1}

Studies and simulations for the performance evaluation of the proposed model were performed by generating data from Eq. \eqref{funG} considering two types of functions $g(\cdot)$: the first formed by a linear combination of B-splines, and the second by a linear combination of trigonometric functions.

Due to the computation time and the various tests that were performed, only five curves were generated ($m=5$) per function for greater flexibility in obtaining the results. Instead of using the raw data directly in the model, it is possible to work with the midpoints of these repeated measures. This approach is computationally more viable and recommended for cases where there is no interest in the individual effect of each curve. Another approach is to work directly with the raw data and the simple mean of the products $Z_{ik}\beta_{ik}$ is calculated with $k$ fixed after obtaining the individual estimates of the partial coefficients and the $Z_{ik}$s. It was decided to work with the second approach in aiming to evaluate the maximum amount of properties existing in the proposed method. It is noteworthy that regardless of the choice, the higher $m$, the lower the aggregate uncertainty will be. Regarding the number of observations per curve, it was established that $n_{i}=n=100, \forall i\in\{1,\dots,m\}$, for both constructed curve scenarios.

Next, a specific grid of equally spaced points for each function was defined without loss of generality, from which the respective functional data were generated. Data with different disturbances were generated in both scenarios of constructed curves, and the following disturbance levels were more specifically considered: $\sigma=0.1$ and $\sigma=0.5$. It is expected that the performance of the models gets better as the noise decreases.

The Bayesian model was implemented through the Gibbs sampler to enable the convergence diagnosis using two chains which start at different points, and with the computation of 10000 iterations. Considering a burn-in period of 50\% of the size of each chain and spacing of 50 points between each sampled value, each chain has 100 sampled points. Thus, a sample size 200 will be obtained at the end of the process for each posterior distribution.

The convergence diagnosis used in this analysis takes into account a chain convergence test proposed by \citeauthor{Gelman} (\citeyear{Gelman}). The test is based on comparing of variances between chains and within these chains, being very similar to the classical analysis of variance. The closer the test statistic is to one, the more evidence we have about the convergence of the chain. For this it is necessary to generate two or more chains in parallel that start at sufficiently distinct points. In the case of the studies and simulations performed herein, the chains were initialized as follows:
\begin{itemize}
	\item $\vec{\beta}=\vec{-1}$, $\vec{\theta}=\vec{\frac{1}{5}}$, $\sigma^2=1$ and $\tau^2=1$ for the first chain and
	\item $\vec{\beta}=\vec{1}$, $\vec{\theta}=\vec{\frac{4}{5}}$, $\sigma^2=5$ and $\tau^2=5$ for the second chain. 
\end{itemize} 

Latent random variables $Z_{ki}$s were initialized in the first chain via a random draw and the complementary vector of the draw was used for the second chain.

Next, the hyperparameters were defined as: $\delta_{1}=\delta_{2}=\psi_{1}=\psi_{2}=0$. Thus, when using a prior for $\sigma^2$ degenerated with $\psi_{1}=0$ and $\psi_{2}=0$, there is an equivalence with the use of the improper non-informative prior $\frac{1}{\sigma^2}$. The interpretation is similar for $\tau^2$.

In situations in which $\vec{\mu}$ is considered as a parameter, it is set $\psi=0.6$ and the first chain was initialized with $\vec{\mu}=\vec{\frac{1}{5}}$, while the second chain was with $\vec{\mu}=\vec{\frac{4}{5}}$. On the other hand, for the cases in which such component was considered a hyperparameter, several models were generated taking into account a grid of nine equally spaced values varying from $0.1$ to $0.9$.

Next, we calculated the MAP (Maximum a Posteriori) estimate for each parameter to summarize the resulting information. Thus, with respect to latent variables $Z_{ki}'$s, we chose the most frequent value, which is the posterior mode.

Finally, unlike the simulations presented in Subsection \ref{secao_replicacoes}, it is important to highlight that the studies with synthetic data performed in Subsections \ref{1s} and \ref{2s} are conducted without replication and, therefore, are not called simulations to avoid confusion. The option to carry out these studies with synthetic data is justified by the computational feasibility. In turn, the main objective is to evaluate the general fitting capacity of the proposed model and study which model configurations among those tested have the potential to generate the best results, since the best configurations obtained herein are later used in the study with replications presented in Subsection \ref{secao_replicacoes}. As there are no replications, the results in Subsections \ref{1s} and \ref{2s} should be observed keeping in mind that they might be slightly different if other datasets were considered.

\subsection{Performance metrics}
\label{MP}
Let $\vec{B}$ be the $n \times K$ matrix of basis function evaluations and $\vec{y_{i.}}=(y_{i1},\;y_{i2},\dots,\;y_{in_{i}})^{'}$ the vector of observed points from the i-th curve. Let $\vec{\hat{\beta}_{k.}}=(\hat{\beta}_{k1},\;\hat{\beta}_{k2},\dots,\;\hat{\beta}_{km})^{'}$ the vector with the MAP estimates of the $\beta_{ki}$'s (fixed $k$) and $\vec{\hat{Z}_{k.}}=(\hat{Z}_{k1},\;\hat{Z}_{k2},\dots,\;\hat{Z}_{km})^{'}$ the vector with the respective most frequent values in the chain for each $Z_{ki}$ (fixed $k$). Thus, for the purpose of comparison and performance studies, we define the global estimate $\vec{\hat{\xi}}=(\hat{\xi}_{1},\;\hat{\xi}_{2},\dots,\;\hat{\xi}_{K})^{'}=(\frac{\vec{\hat{Z}_{1.}}'\vec{\hat{\beta}_{1.}}}{m},\;\frac{\vec{\hat{Z}_{2.}}'\vec{\hat{\beta}_{2.}}}{m},\dots,\;\frac{\vec{\hat{Z}_{K.}}'\vec{\hat{\beta}_{K.}}}{m})^{'}$ as being the vector with the final average estimates of the coefficients from the linear combination of the basis functions. Therefore, we can write

\begin{equation}
	1-\frac{1}{m}\sum_{i=1}^{m}	
	\frac{(n_{i}-1)(\vec{y_{i.}}-\vec{B}\vec{\hat{\xi}})^{'}(\vec{y_{i.}}-\vec{B}\vec{\hat{\xi}})}{\left(n_{i}-\sum_{k=1}^{K}I_{\left\{|\hat{\xi}_{k}|>0\right\}}\right)\left(\vec{y_{i.}}-\frac{1}{n_{i}}\sum_{j=1}^{n_{i}}y_{ij}\right)^{'}\left(\vec{y_{i.}}-\frac{1}{n_{i}}\sum_{j=1}^{n_{i}}y_{ij}\right)}
	\label{eq:r2}
\end{equation}as a metric similar to the adjusted $R^{2}$, but it is restricted to the final number of selected basis functions,  $K_{\text{end}}=\sum_{k=1}^{K}I_{\left\{|\hat{\xi}_{k}|>0\right\}}$. Thus, the closer to one this metric is, the better the performance. Note that the indicator function in $K_{\text{end}}$ takes zero value only if all final estimates of the $k$th coefficient are zero. 


In addition to the metric presented in Expression \eqref{eq:r2}, comparisons and performance evaluations of results were also performed using the Mean Squared Error (MSE). The use of the MSE was restricted to the individual assessment of the quality of the model’s fit, meaning for the assessment of the proximity between the estimated and the true curve, while the metric presented in Expression \eqref{eq:r2} is used for the comparison between the models since it imposes a penalty on models with more parameters.


In the studies with synthetic data presented in Sections 3.3.1 and 3.3.2, the performance of the proposed models is also compared with the ordinary least squares (OLS) fit, for which no regularization is imposed.

\subsection{Results}
\subsubsection{\texorpdfstring{$1^{\text{st}}$}{} Study with synthetic data: linear combination of B-splines}
\label{1s}

In this section, we evaluate both the potential that the proposed model has to perform the selection of bases as well as its ability to select them correctly whenever the real bases that generated the curve are found within the set of candidate bases. Therefore, we generated functional data arising from a linear combination of B-splines with known true coefficients plus some random noise. Data were generated considering the same set of evaluation points, $t_{ij}$'s, defined by the function \verb|seq(0, 1, length=100)| of the R software program. Regarding the noise variance, $\sigma^2=(0.1)^2$ was considered to generate five curves, and $\sigma^2=(0.5)^2$ to generate another set of five curves. Four coefficients were purposely set to zero. Thus, the vector of true coefficients is $(-2,0,1.5,1.5,0,-1,-0.5,-1,0,0)$, so that:
\begin{equation}
y({t_{ij}})= (-2,0,1.5,1.5,0,-1,-0.5,-1,0,0)^{'}\begin{pmatrix}
\mathcal{B}_{1}({t_{ij}})\\\mathcal{B}_{2}({t_{ij}})\\\vdots\\\mathcal{B}_{10 }({t_{ij}})
\end{pmatrix}+\epsilon_{ij}\text{, with }\epsilon_{ij}\sim\normal(0,\sigma^2)\text{,}
\label{1função}
\end{equation} where $(\mathcal{B}_{1}({t_{ij}}),\;\mathcal{B}_{2}({t_{ij}}),\dots, \;\mathcal{B}_{10}({t_{ij}}))^{'}$ is the vector with the B-splines evaluated at $t_{ij}$ used to generate the $j$th point of the $i$th synthetic curve. 

It is expected that a satisfactory base selection model will not only be able to estimate all the coefficients well, but also that it will be able to exclude all three bases associated with the null coefficients from the model.

Assuming that in practice researchers have no information regarding to which set of basis functions is ideal, tests were carried out with B-splines considering 14 possibilities for the total number of bases ($K \in \{5,6,7,8,9,10,11,12,13,14,15,20,25,30\}$). The search grid for $K$ is more refined in the neighborhood of 10, which is the value purposely used to generate the data in order to assess the model’s sensitivity.

It is noteworthy that the diagnostic analysis based on the method proposed by \citeauthor{Gelman} (\citeyear{Gelman}) attested to the convergence of the partial coefficient chains after the burn-in period in all tested model configurations.

\begin{table}[!htb]
	\centering
	{\fontsize{4.5}{11}\selectfont 
		\rotatebox{0}{
\begin{tabular}{cc|ccccccccc|cc}\hline
	&  & \multicolumn{9}{c|}{$\mu$ (Hyperparameter)} &  &  \\\cline{3-11}
	\multirow{-2}{*}{$\sigma$} & \multirow{-2}{*}{$K$} & 0.1 & 0.2 & 0.3 & 0.4 & 0.5 & 0.6 & 0.7 & 0.8 & 0.9 & \multirow{-2}{*}{$\mu$ (Parameter)} & \multirow{-2}{*}{OLS} \\\hline
	& 5 & 0.97083 & 0.97083 & 0.97082 & 0.97083 & 0.97083 & \cellcolor[HTML]{6699FF}0.97086 & 0.97077 & 0.97084 & 0.97070 & 0.97085 & 0.97096  \\
	& 6 & 0.96837 & 0.96837 & 0.96834 & \cellcolor[HTML]{6699FF}0.96839 & 0.96830 & 0.96827 & 0.96836 & 0.96836 & 0.96833 & 0.96832 & 0.96811 \\
	& 7 & 0.97955 & 0.97952 & 0.97989 & 0.97985 & 0.97994 & 0.97994 & 0.97994 & 0.97995 & \cellcolor[HTML]{6699FF}0.97998 & 0.97957 & 0.98002  \\
	& 8 & 0.97603 & 0.97608 & 0.97633 & 0.97618 & 0.97646 & 0.97644 & 0.97653 & 0.97662 & \cellcolor[HTML]{6699FF}0.97680 & 0.97575 & 0.97686 \\
	& 9 & 0.97994 & 0.98004 & 0.98050 & 0.98095 & 0.98049 & 0.98108 & 0.98111 & 0.98126 & \cellcolor[HTML]{6699FF}0.98131 & 0.97984 & 0.98140	 \\
	& {\color[HTML]{FF0000} 10} & \cellcolor[HTML]{6699FF}{\color[HTML]{FF0000} 0.98359} & 0.98358 & 0.98358 & 0.98338 & 0.98340 & 0.98341 & 0.98329 & 0.98331 & 0.98315 & {\color[HTML]{FF0000} 0.98357} & {\color[HTML]{FF0000} 0.98298} \\
	& 11 & 0.98217 & 0.98257 & 0.98255 & 0.98257 & 0.98247 & \cellcolor[HTML]{6699FF}0.98264 & 0.98254 & 0.98249 & 0.98251 & 0.98225 & 0.98260 \\
	& 12 & 0.98280 & \cellcolor[HTML]{6699FF}0.98286 & 0.98278 & 0.98274 & 0.98277 & 0.98264 & 0.98272 & 0.98258 & 0.98263 & 0.98272 & 0.98269 \\
	& 13 & 0.98287 & 0.98287 & \cellcolor[HTML]{6699FF}0.98290 & 0.98285 & 0.98287 & 0.98275 & 0.98268 & 0.98252 & 0.98258 & 0.98285 & 0.98265 \\
	& 14 & 0.98262 & 0.98259 & \cellcolor[HTML]{6699FF}0.98278 & 0.98277 & 0.98275 & 0.98269 & 0.98268 & 0.98272 & 0.98246 & 0.98262 & 0.98257	\\
	& 15 & 0.98197 & 0.98228 & 0.98227 & 0.98226 & \cellcolor[HTML]{6699FF}0.98242 & 0.98227 & 0.98218 & 0.98221 & 0.98225 & 0.98195 & 0.98233 \\
	& 20 & 0.98028 & 0.98096 & 0.98107 & 0.98110 & 0.98136 & \cellcolor[HTML]{6699FF}0.98139 & 0.98125 & 0.98122 & 0.98127 & 0.98031 & 0.98144 \\
	& 25 & 0.97977 & 0.98047 & 0.98061 & \cellcolor[HTML]{6699FF}0.98081 & 0.98048 & 0.98072 & 0.98037 & 0.98025 & 0.98020 & 0.97980 & 0.98050 \\
	\multirow{-14}{*}{0.1} & 30 & 0.97602 & 0.97891 & \cellcolor[HTML]{6699FF}0.97998 & 0.97986 & 0.97967 & 0.97962 & 0.97949 & 0.97913 & 0.97930 & 0.97526 & 0.97944  \\\hline
	& 5 & 0.68136 & 0.68161 & 0.68118 & 0.68102 & 0.68175 & \cellcolor[HTML]{6699FF}0.68488 & 0.68449 & 0.68072 & 0.68201 & 0.68136 & 0.68251 \\
	& 6 & 0.67527 & 0.68003 & 0.68183 & 0.68445 & 0.68396 & \cellcolor[HTML]{6699FF}0.68466 & 0.68167 & 0.68286 & 0.68364 & 0.67777 & 0.68418 \\
	& 7 & 0.68046 & 0.68036 & 0.68489 & 0.68611 & 0.68799 & 0.68708 & 0.68651 & 0.68686 & \cellcolor[HTML]{6699FF}0.68887 & 0.68239 & {\color[HTML]{FF0000} 0.69018}		 \\
	& 8 & 0.68254 & 0.68339 & 0.68265 & 0.68288 & 0.68040 & 0.67996 & 0.68244 & 0.68025 & \cellcolor[HTML]{6699FF}0.68422 & 0.68166 & 0.68538 \\
	& 9 & 0.67998 & 0.68067 & 0.68084 & 0.67675 & 0.68040 & 0.68175 & 0.68257 & 0.68218 & \cellcolor[HTML]{6699FF}0.68259 & 0.67949 & 0.68427 \\
	& {\color[HTML]{FF0000} 10} & \cellcolor[HTML]{6699FF}{\color[HTML]{FF0000} 0.69242} & 0.68751 & 0.68860 & 0.68672 & 0.68667 & 0.68425 & 0.68139 & 0.68045 & 0.68090 & {\color[HTML]{FF0000} 0.69180} &0.68285  \\
	& 11 & 0.68317 & 0.68567 & \cellcolor[HTML]{6699FF}0.68829 & 0.68596 & 0.68663 & 0.68657 & 0.68203 & 0.67957 & 0.67988 & 0.68404 & 0.68087 \\
	& 12 & 0.68023 & 0.68069 & 0.68277 & \cellcolor[HTML]{6699FF}0.68336 & 0.68326 & 0.67974 & 0.67617 & 0.67863 & 0.67887 & 0.67917 & 0.68093	 \\
	& 13 & \cellcolor[HTML]{6699FF}0.68470 & 0.67951 & 0.68002 & 0.67744 & 0.67789 & 0.67922 & 0.67376 & 0.67382 & 0.67456 & 0.68363 & 0.67686	 \\
	& 14 & 0.68254 & \cellcolor[HTML]{6699FF}0.68294 & 0.68170 & 0.67942 & 0.67152 & 0.67293 & 0.67437 & 0.67163 & 0.66957 & 0.68351 &  0.67360\\
	& 15 & 0.67668 & 0.67613 & \cellcolor[HTML]{6699FF}0.67687 & 0.67297 & 0.66893 & 0.66786 & 0.66803 & 0.66811 & 0.66807 & 0.67620 & 0.67070 \\
	& 20 & 0.63714 & 0.64613 & 0.65175 & 0.65237 & \cellcolor[HTML]{6699FF}0.65255 & 0.65006 & 0.64706 & 0.64978 & 0.64904 & 0.64308 & 0.65372 \\
	& 25 & 0.60181 & 0.63506 & 0.63947 & \cellcolor[HTML]{6699FF}0.64654 & 0.64064 & 0.63123 & 0.63136 & 0.63147 & 0.63012 & 0.59338 & 0.63691 \\
	\multirow{-14}{*}{0.5} & 30 & 0.60454 & 0.62355 & \cellcolor[HTML]{6699FF}0.63227 & 0.61635 & 0.61323 & 0.60961 & 0.61127 & 0.61089 & 0.61029 & 0.60124 & 0.61659 \\\hline
\end{tabular}
		}
		\caption{\textit{$1^{\text{st}}$ study with synthetic data}. Performance metrics (Eq. \ref{eq:r2}) of the proposed Bayesian approach with $\mu$ as hyperparameter and as parameter according to the different configurations of $K$ bases and two noise variances, $\sigma^2 = (0.1)^2$ and $\sigma^2 = (0.5)^2$. Results for the OLS method is also presented as a comparison.}
		\label{tab:my-table}
   }
\end{table}

Table \ref{tab:my-table} presents the performance metrics calculated in accordance with Eq. \eqref{eq:r2} for the different model configurations tested, in addition to presenting the performance of the Ordinary Least Squares (OLS) method as a function of the total number of bases. As five functionals are being used and the OLS method does not estimate jointly the coefficients of each functional, then the OLS method was applied to each of the five functionals and subsequently the simple mean of the estimated coefficients was used for the computation of performance metrics. The cells highlighted in the table indicate the best model in each line among those with $\mu$ considered as a hyperparameter. 

A brief analysis of the results presented in Table \ref{tab:my-table} shows that the proposed method is fully capable of detecting the best $K$ to be used. The results show that the best model is the one with $K=10$ and $\mu=0.1$, for both $\sigma=0.1$ and $\sigma=0.5$, which is in agreement with the $K$ used in generating the synthetic data. The same does not happen with the Ordinary Least Squares method, which although it manages to point to $K=10$ in the study with small sigma, it does not have the same capacity to indicate the best $K$ in the study with large sigma. It is interesting to note that the models whose configuration takes into account $\mu$ as a parameter had a satisfactory performance as the others with $\mu$ as a hyperparameter.

Supplementary Material Table 1 presents the MSEs of the respective models, which are calculated with the information from the estimated curve and the theoretical curve, and shows the great capacity that the proposed model has to generate estimated curves close to the true one. The best model in terms of MSE for small variance presents an MSE to the order of ten to minus five, while this measure for large variance is to the order of ten to minus three. Naturally, in the study with large sigma, the model’s ability to explain the total variability of the data is smaller than in the study with small sigma, suggesting that the greater the variance of the data, the worse the fit.

Regarding the MSEs returned by OLS methods, it is possible to observe in Supplementary Material Table 1, for $K=10$ and $\sigma=0.1$, a substantial difference between the MSE of the best configuration of the proposed model and the MSE obtained by the traditional OLS method. While the first is in the order of ten to minus five, the other is in the order of ten to minus four. This discrepancy is not as significant when such comparison is made, taking into account $\sigma=0.5$, although the OLS method still returns a higher MSE.
 
Another relevant aspect to be highlighted is the ability of the OLS method to indicate the best $K$ when using MSE for both choices of $\sigma$ (Supplementary Material Table 1). However, as described earlier, the OLS method for $\sigma=0.5$ is not able to point the true $K$ through the metric \eqref{eq:r2} (Table \ref{tab:my-table}). This is largely justified by the very construction of the metric \eqref{eq:r2}, which penalizes the excess of bases selected by the model. As the OLS method is not able to perform variable selection, it will be greatly penalized by the metric \eqref{eq:r2}. Thus, the value of the metric \eqref{eq:r2} for OLS with $K=10$ ends up being lower than that returned by the referred model when considering $K=7$, which is a similar configuration to the one with $K=10$, but which already imposes a smaller amount of parameters.

Table \ref{tab:coeficientes-curvaBS} presents the information needed to compare the parameters of the linear combination coefficients and their respective estimates obtained by the best models according to the metric in Eq. \eqref{eq:r2}. As five functionals are used in the study, each estimate in the table is calculated through the simple average of the respective final estimates of the coefficients of each functional, meaning the final average estimate of the $k$th coefficient is given by $\hat{\xi}_{k}=\frac{\vec{\hat{Z}_{k.}}'\vec{\hat{\beta}_{k.}}}{m}$, as described in Section \ref{MP}.

\begin{table}[!htb]
	{\fontsize{7.5}{14}\selectfont 
		\begin{tabular}{c|cccccccccc}\hline
		\multirow{2}{*}{} & \multicolumn{10}{c}{Final average estimates} \\\cline{2-11}
				& $\hat{\xi}_1$ & $\hat{\xi}_2$ & $\hat{\xi}_3$ & $\hat{\xi}_4$ & $\hat{\xi}_5$ & $\hat{\xi}_6$ & $\hat{\xi}_7$ & $\hat{\xi}_8$ & $\hat{\xi}_9$ & $\hat{\xi}_{10}$ \\\hline
			With $\sigma=0.1$&-2.01733&0&1.50971&1.47201&0&-1.00098&-0.48140&-0.99795&0&0\\
			With $\sigma=0.5$ &-1.98899&0&1.53058&1.37556&0&-1.06762&-0.25667&-1.01188&0&0\\
			True coefficient& -2 & 0 & 1.5 & 1.5 & 0 & -1 & -0.5 & -1 & 0 & 0\\\hline
		\end{tabular}
		\caption{Comparative table of the final average estimates with the true coefficients, according to the data dispersion degree.}
		\label{tab:coeficientes-curvaBS}
	}
\end{table}

Regardless of the variance of the synthetic data, it is clear that the model is not only capable of promoting the selection of bases, but is also capable of exactly selecting the bases that were used in the data generation. It is noteworthy that the estimates were sufficiently close to the true ones.

We should also highlight that there is a perfect agreement on which bases are selected among the final individual functional estimates; since if at least one functional disagreed with the others on the value of the latent variable for a given coefficient, the final average estimate would be different from zero.

\begin{figure}[!htb]
	\centering
	\subfloat[Data with $\sigma=0.1$ ($K=10$ and $\mu=0.1$)\label{bs1:01}]{\includegraphics[scale=0.34]{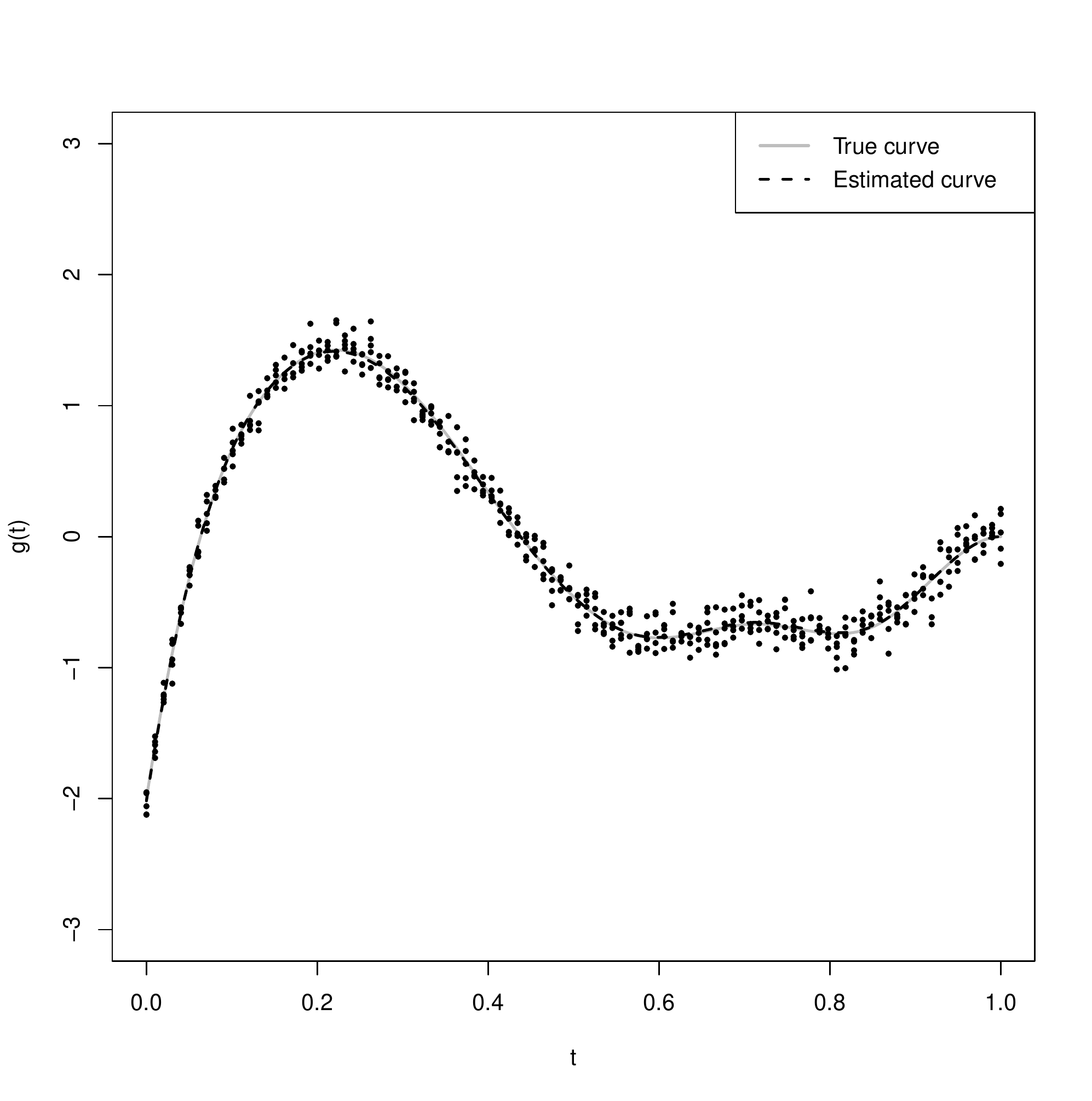}}
	\subfloat[Data with $\sigma=0.5$ ($K=10$ and $\mu=0.1$)\label{bs1:05}]{\includegraphics[scale=0.34]{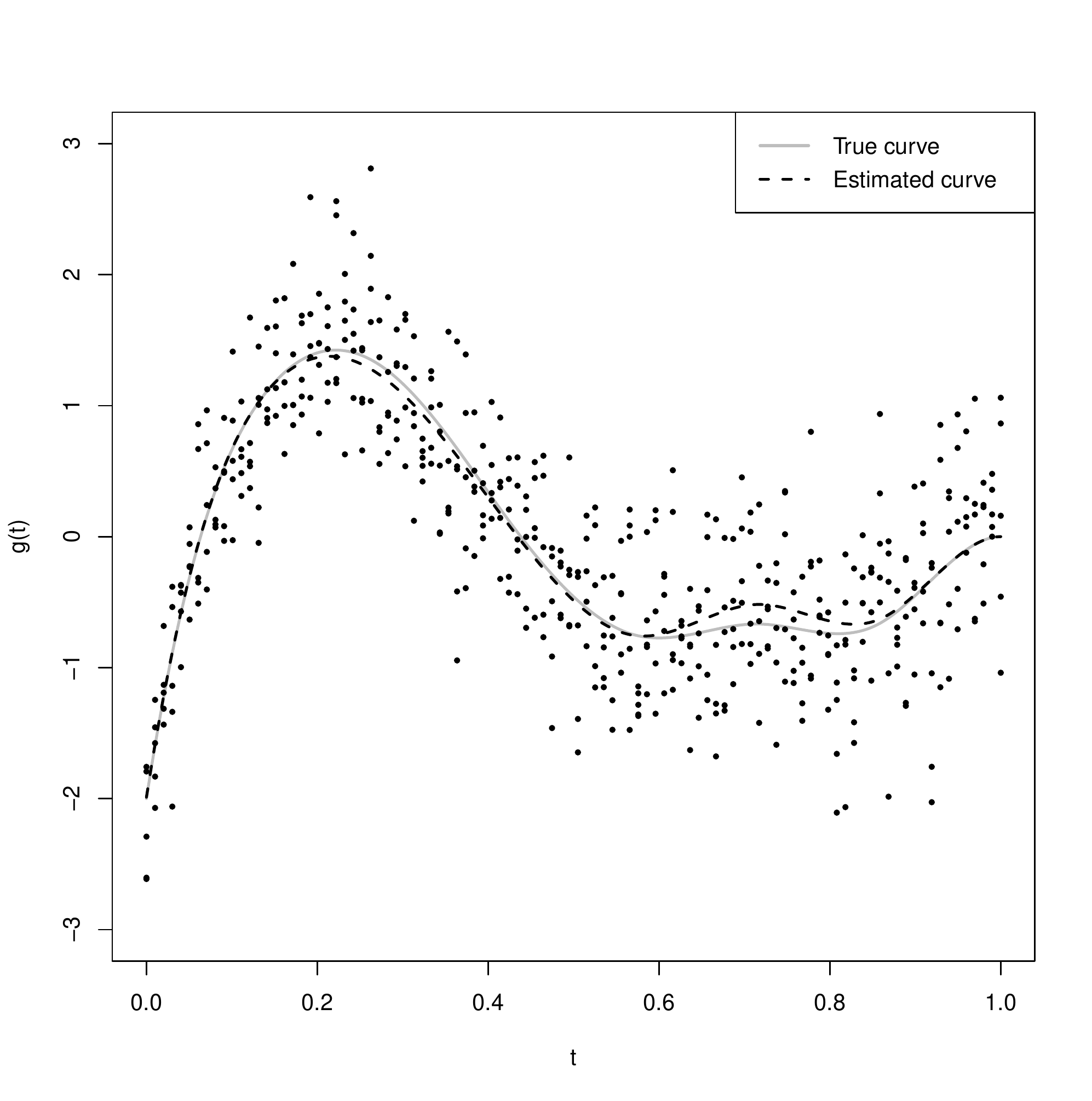}}
	\caption{Data ($y_{t_{ij}}$) and curves (true and estimated) by the proposed model with the B-splines, according to the data dispersion degree ($1^{\text{st}}$ study with synthetic data).}
	\label{fig:bs1}
\end{figure}

Finally, Figure \ref{fig:bs1} presents the estimated function $g(\cdot)$ by the best models according to the metric in Eq. \eqref{eq:r2} for each of the respective error standard deviations used in this study. Just by graphical visualization, it is not possible to notice a difference between the true and estimated functions in Figure \ref{bs1:01}, while it is already possible to see a difference in the interval between $0.6$ and $0.8$ between the true and estimated curves in Figure \ref{bs1:05}, which is a region greatly influenced by the seventh coefficient.

\subsubsection{\texorpdfstring{$2^{\text{nd}}$}{} study with synthetic data: linear combination of trigonometric functions}
\label{2s}

For this second study we generate data according to:
\begin{equation}
	y(t_{ij})=\cos(t_{ij})+\sin(2t_{ij})+\epsilon_{ij}\text{, with}\quad\epsilon_{ij}\sim\normal(0,\sigma^2)\text{.}
	\label{eq:2}
\end{equation}

\noindent where the evaluation points, $t_{ij}$'s, are defined by the R function \verb|seq(0, 2*pi, length=100)|. As in the previous study, five curves were generated with $\sigma^2=(0.1)^2$ and 5 others with $\sigma^2=(0.5)^2$. As the data are generated from trigonometric functions which induce periodicity, the model is not only be tested with B-splines bases, but also with Fourier bases.

Due to the orthogonality of Fourier bases (\citeauthor{zygmund2002trigonometric}; \citeyear{zygmund2002trigonometric}), the procedure previously used in Section 3.3.1 to choose the maximum number of bases, $K$, is not necessary. In other words, using Fourier bases is enough to fix a $K$ large enough for the proposed model to be able to select the ideal bases for the adjustment. Therefore, $K=30$ was defined for the model with Fourier bases. As there is no orthogonality between the bases for the B-splines, the choice of $K$ influences the fit quality. Thus, tests were performed considering six possibilities for the total number of basis functions ($K\in\{5,10,15,20,25,30\}$).

The diagnostic analysis in this second study based on the method proposed by \citeauthor{Gelman} (\citeyear{Gelman}) also attested to the convergence of the partial coefficient chains after the burn-in period in all tested model configurations.

Table \ref{tab:my-table2} presents the performance metrics calculated in accordance with Expression \eqref{eq:r2} for the various model configurations tested, as well as the performance of the OLS method as a function of the total number of B-spline basis functions.

\begin{table}[!htb]
	\centering
	{\fontsize{4.5}{11}\selectfont 
		\rotatebox{0}{
\begin{tabular}{cc|ccccccccc|cc}\hline
	&  & \multicolumn{9}{|c|}{$\mu$ (Hyperparameter)} &  &  \\\cline{3-11}
	\multirow{-2}{*}{$\sigma$} & \multirow{-2}{*}{K} & 0.1 & 0.2 & 0.3 & 0.4 & 0.5 & 0.6 & 0.7 & 0.8 & 0.9 & \multirow{-2}{*}{$\mu$ (Parameter)} & \multirow{-2}{*}{OLS} \\\hline
	& 5 & 0.59853 & 0.59895 & 0.59849 & 0.59898 & 0.59672 & 0.59823 & 0.59645 & \cellcolor[HTML]{6699FF}0.60280 & 0.60183 & 0.59874 & 0.60556 \\
	& 10 & 0.98791 & \cellcolor[HTML]{6699FF}0.98793 & 0.98791 & 0.98787 & 0.98786 & 0.98789 & 0.98793 & 0.98792 & 0.98792 & 0.98792 & 	0.98794	\\
	& {\color[HTML]{FF0000} 15} & 0.98812 & 0.98809 & \cellcolor[HTML]{6699FF}{\color[HTML]{FF0000} 0.98814} & 0.98800 & 0.98804 & 0.98798 & 0.98796 & 0.98805 & 0.98807 & {\color[HTML]{FF0000} 0.98805} &{\color[HTML]{FF0000} 0.98813}   \\
	& 20 & 0.98693 & 0.98690 & 0.98693 & 0.98707 & 0.98727 & 0.98727 & 0.98737 & 0.98738 & \cellcolor[HTML]{6699FF}0.98742 & 0.98691 & 0.98753 \\
	& 25 & 0.98585 & 0.98618 & 0.98650 & 0.98661 & 0.98674 & 0.98665 & 0.98676 & 0.98673 & \cellcolor[HTML]{6699FF}0.98683 & 0.98594 & 0.98692 \\
	\multirow{-6}{*}{0.1} & 30 & 0.98573 & 0.98588 & 0.98611 & 0.98606 & 0.98630 & 0.98627 & 0.98626 & \cellcolor[HTML]{6699FF}0.98639 & 0.98611 & 0.98587 & 0.98621 \\\hline
	& 5 & 0.44966 & 0.44808 & 0.45146 & \cellcolor[HTML]{6699FF}0.45611 & 0.45198 & 0.45202 & 0.44624 & 0.45041 & 0.45296 & 0.43827 & 0.45860	 \\
	& {\color[HTML]{FF0000} 10} & 0.76060 & 0.76424 & 0.76291 & 0.77013 & 0.77151 & 0.77029 & 0.76743 & 0.76967 & \cellcolor[HTML]{6699FF}{\color[HTML]{FF0000} 0.77328} & {\color[HTML]{FF0000} 0.75910} & {\color[HTML]{FF0000}0.77416} \\
	& 15 & 0.74816 & 0.75369 & 0.76072 & 0.76397 & 0.76358 & 0.76485 & 0.76607 & \cellcolor[HTML]{6699FF}0.76617 & 0.76595 & 0.74735 & 0.76705 \\
	& 20 & 0.73483 & 0.74817 & 0.75575 & 0.75237 & \cellcolor[HTML]{6699FF}0.75722 & 0.75480 & 0.75338 & 0.75340 & 0.75198 & 0.73218 & 0.75518 \\
	& 25 & 0.71552 & 0.71733 & 0.73931 & 0.73891 & 0.73723 & 0.74203 & \cellcolor[HTML]{6699FF}0.74437 & 0.73847 & 0.74110 & 0.70228 & 0.74313 \\
	\multirow{-6}{*}{0.5} & 30 & 0.69127 & 0.71311 & \cellcolor[HTML]{6699FF}0.72995 & 0.72798 & 0.72100 & 0.72089 & 0.72401 & 0.72526 & 0.72600 & 0.69042 & 0.72966 \\\hline
\end{tabular}
		}
		\caption{\textit{$2^{\text{nd}}$ study with synthetic data}. Performance metrics (Eq. \ref{eq:r2}) of the proposed Bayesian approach with $\mu$ as hyperparameter and as parameter according to the different configurations of $K$ B-spline basis functions and two noise variances, $\sigma^2 = (0.1)^2$ and $\sigma^2 = (0.5)^2$. Results for the OLS method is also presented as a comparison.}
		\label{tab:my-table2}
	}
\end{table}

The results point to good performance by the model in several configurations, showing its ability to perform good fits even when the real bases are not found in the set of candidate bases. However, as a consequence of the use of candidate bases which are not part of construction of true mean function, the procedure adopted for choosing the best $K$ suggests different values depending on the data dispersion degree, unlike what occurred in the study presented in Section 3.3.1, since $K=15$ was indicated for data with small variance, while the model with the best fit for data with greater dispersion was with 
$K=10$.

In this study, the discrepancy between the performance of models with $\mu$ as a parameter and with $\mu$ as a hyperparameter is greater than in the previous study. Naturally, the automated version of the proposed model, meaning the one with $\mu$ as a parameter, is more appropriate to situations in which there is already a prior knowledge of the need to exclude some bases due to the limitation already described above, which requires a uniform prior with support between 0 and $\psi$, where $\psi$ cannot be so close to 1.

Supplementary Material Table 2 presents the MSEs comparing the true and estimated mean functions from the respective models. In addition, the results in this table confirm the previous analysis regarding the absence of a consensus for choosing the best $K$.

\begin{table}[!htb]
\centering
{\fontsize{6}{17}\selectfont 
\begin{tabular}{c|cc}\hline
\multirow{2}{*}{Final average estimates}&\multicolumn{2}{c}{$\sigma$}\\\cline{2-3}
       & 0.1      & 0.5      \\ \hline
 $\hat{\xi}_1$ & 0.95738  & 0.87897 \\  
 $\hat{\xi}_2$    & 1.41036  & 1.57066  \\ 
 $\hat{\xi}_3$    & 1.87265  & 2.32558  \\ 
 $\hat{\xi}_4$    & 1.58369  & -1.06712 \\ 
 $\hat{\xi}_5$    & 0        & -2.44673 \\ 
 $\hat{\xi}_6$    & -1.62446 & 0.42485  \\ 
 $\hat{\xi}_7$    & -1.92588 & 0.57600  \\ 
 $\hat{\xi}_8$    & -1.03001 & -0.94346 \\ 
 $\hat{\xi}_9$    & 0.10636  & 0.43142  \\ 
 $\hat{\xi}_{10}$ & 0.55217  & 1.17847  \\ 
 $\hat{\xi}_{11}$ & 0        & -        \\ 
 $\hat{\xi}_{12}$ & -0.57727 & -        \\ 
 $\hat{\xi}_{13}$ & 0        & -        \\ 
 $\hat{\xi}_{14}$ & 0.59828  & -        \\ 
 $\hat{\xi}_{15}$ & 1.05177  & -        \\\hline 
\end{tabular}
\caption{Final average estimates of the coefficients of the model implemented with B-splines bases according to the data dispersion degree.}
\label{tab:xiBs}
}
\end{table}

Table \ref{tab:xiBs} presents the final average estimates obtained for the best $K$ under each data dispersion scenario considered for better understanding of which and how many basis functions were excluded from the model. It is important to remember that the set of bases used in the data with $\sigma=0.5$ is not contained in the set of bases defined for the data with $\sigma=0.1$, therefore, although the bases in both scenarios are B-splines, such bases are different, and it is not possible to make comparisons between the coefficients of each scenario.

\begin{figure}[!htb]
	\centering
	\subfloat[Data with  $\sigma=0.1$ ($K=15$ and  $\mu=0.3$)\label{bs2:01}]{\includegraphics[scale=0.34]{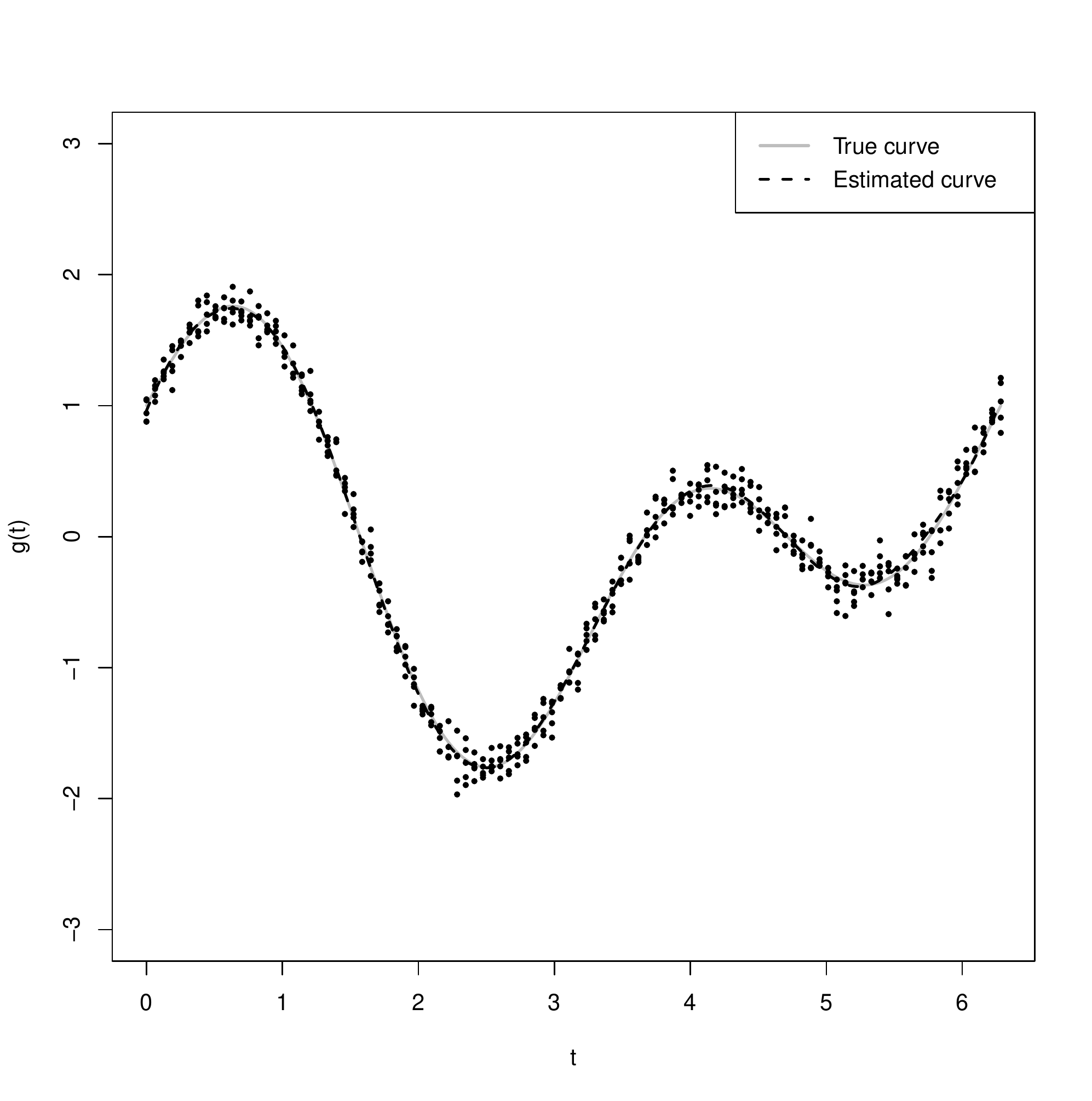}}
	\subfloat[Data with  $\sigma=0.5$ ($K=10$ and  $\mu=0.9$)\label{bs2:05}]{\includegraphics[scale=0.34]{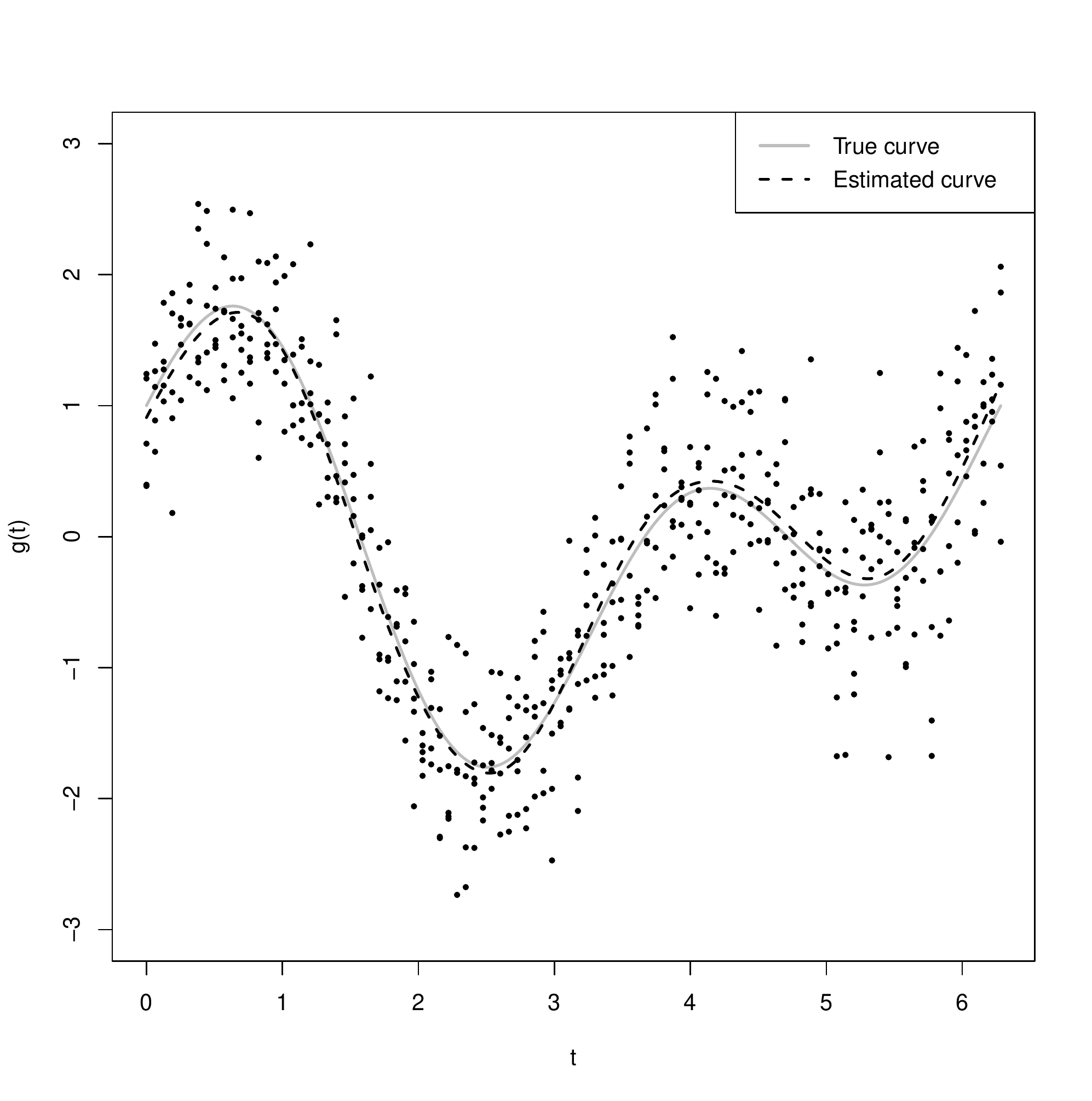}}
	\caption{Data ($y_{t_{ij}}$) and curves (true and estimated) by the proposed model with the B-splines according to the data dispersion degree ($2^{\text{nd}}$ study with synthetic data).}
	\label{fig:bs2}
\end{figure}

Figure \ref{fig:bs2} presents the true curves and the ones estimated by the best models according to the study metrics to conclude the analysis of the models implemented with B-spline bases. Even with the B-spline bases being used in the model construction, which are bases that were not used in the generation of these synthetic data, and even in the study with high dispersion data, the proposed model was able to provide estimated curves that capture the functional curve sign.

As mentioned before, models were also fitted with Fourier bases, which are more suited to problems with periodic curves. Table \ref{tab:F} and Supplementary Material Table 3 present all the results associated with the models implemented with Fourier bases. As the difference between the metrics was observed in the seventh decimal place in some cases, the best models are presented in the table with their respective cells highlighted.

	
	{\centering\fontsize{6}{15}\selectfont 
		\begin{longtable}[c]{c|cc}
	\caption{Performance metrics \eqref{eq:r2} of the OLS methods and the proposed Bayesian models according to the different configurations tested with Fourier bases ($2^{\text{nd}}$ study with synthetic data).}
    \label{tab:F}\endfirsthead\endhead\hline
				\multicolumn{1}{c|}{} & \multicolumn{2}{c}{$\sigma$} \\\cline{2-3}
				\multicolumn{1}{c|}{\multirow{-2}{*}{$\mu$ (Hyperparameter)}} & \multicolumn{1}{l}{0.1} & \multicolumn{1}{l}{0.5} \\\hline
				0.01 & 0.98941 & \cellcolor[HTML]{6699FF}0.79227 \\
				0.02 & 0.98940 & 0.79144 \\
				0.03 & 0.98941 & 0.79135 \\
				0.04 & \cellcolor[HTML]{6699FF}0.98941 & 0.79142 \\
				0.05 & 0.98941 & 0.79125 \\
				0.06 & 0.98941 & 0.79144 \\
				0.07 & 0.98941 & 0.79141 \\
				0.08 & 0.98941 & 0.79144 \\
				0.09 & 0.98941 & 0.79143 \\
				0.1 & 0.98936 & 0.79129 \\
				0.2 & 0.98936 & 0.78971 \\
				0.3 & 0.98936 & 0.78864 \\
				0.4 & 0.98936 & 0.78238 \\
				0.5 & 0.98936 & 0.77594 \\
				0.6 & 0.98934 & 0.76606 \\
				0.7 & 0.98914 & 0.74772 \\
				0.8 & 0.98860 & 0.72868 \\
				0.9 & 0.98759 & 0.72452 \\\hline
				\multicolumn{1}{c|}{$\mu$ (Parameter)} & 0.98936 & 0.79144 \\
				\multicolumn{1}{c|}{OLS} &0.98616	&0.72856 \\\hline
			\end{longtable}
	}

The OLS method is evidently not capable of selecting variables and this has a direct impact on its performance, since all $K=30$ Fourier bases are used. Unlike the models fitted with B-spline bases, the models fitted using Fourier bases (Table \ref{tab:F}) exclude practically all bases, selecting only two among the 30 candidates. Table \ref{tab:xiF} shows the final non-zero average estimates for the best Fourier basis fit results. This selection of only two Fourier bases happens for both choices of $\sigma$, and is an expected result since the true curve is formed by a linear combination of two trigonometric functions.

\begin{table}[!htb]
	\centering
	{\centering\fontsize{9}{16}\selectfont 
		\begin{tabular}{c|cc}\hline
			\multirow{2}{*}{Final non-zero average estimates}& \multicolumn{2}{c}{$\sigma$} \\\cline{2-3}
			& 0.1 & 0.5 \\\hline
			$\hat{\xi}_{2}$ & 1.77794 & 1.810486 \\
			$\hat{\xi}_{3}$ & 1.78188 & 1.811544\\\hline
		\end{tabular}
		\caption{Final non-zero average coefficient estimates of the model implemented with Fourier bases, according to the data dispersion degree.}
		\label{tab:xiF}
	}
\end{table}

\begin{figure}[!htb]
	\centering
	\subfloat[Data with  $\sigma=0.1$ ($\mu=0.04$)\label{f2:01}]{\includegraphics[scale=0.34]{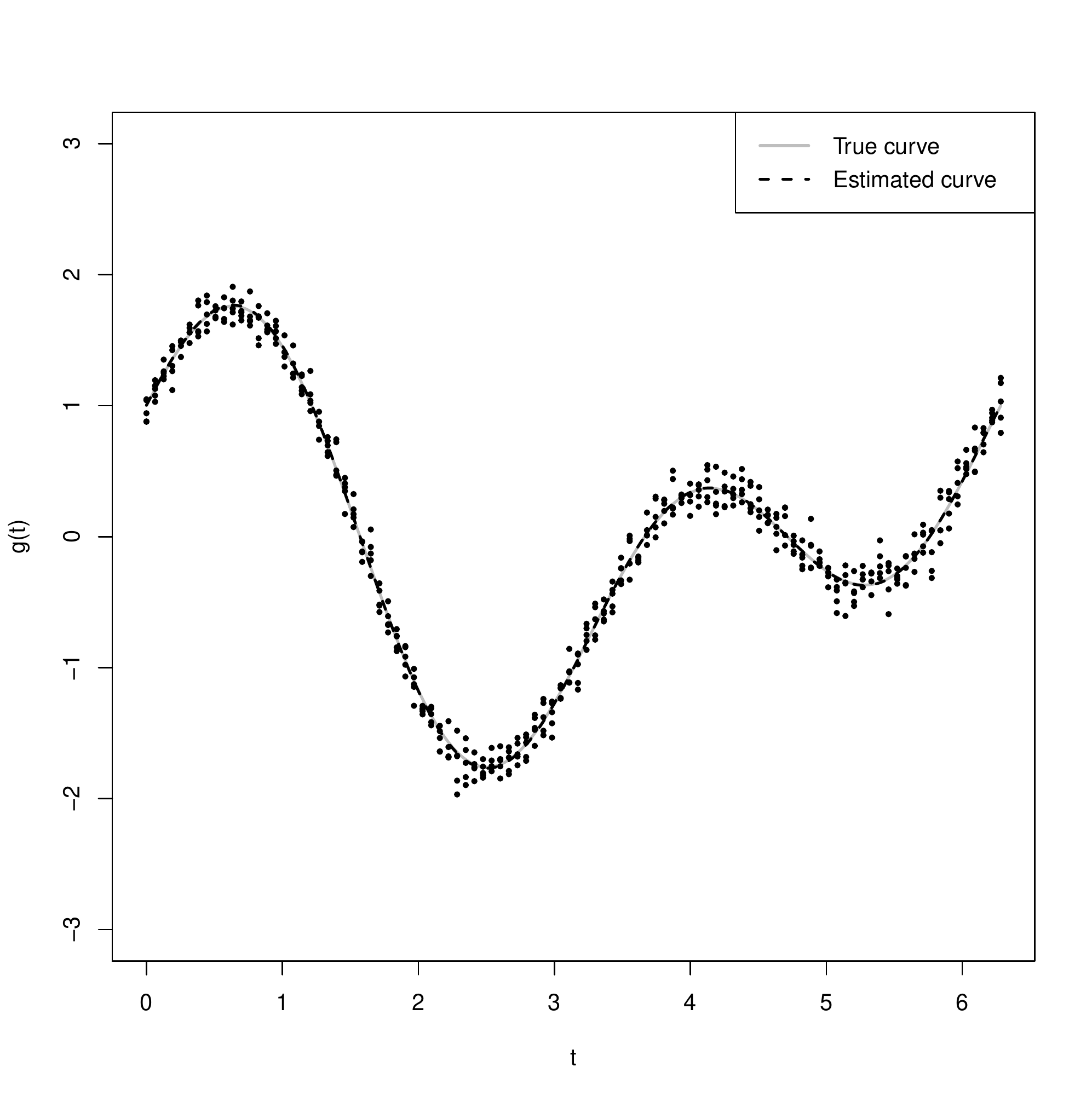}}
	\subfloat[Data with  $\sigma=0.5$ ($\mu=0.01$)\label{f2:05}]{\includegraphics[scale=0.34]{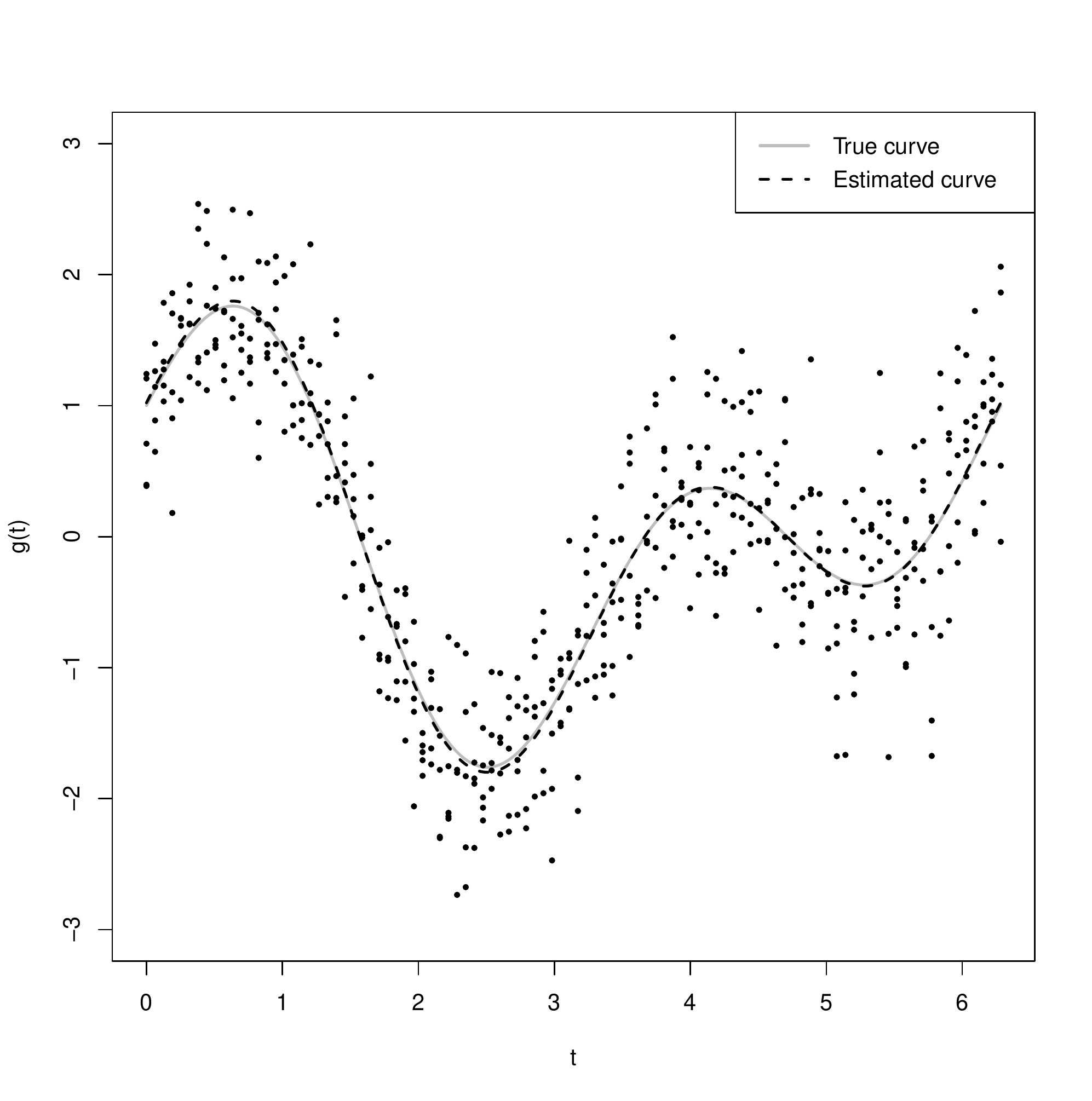}}
	\caption{Data ($y_{t_{ij}}$) and curves (true and estimated) by the proposed model with Fourier bases according to the data dispersion degree ($2^{\text{nd}}$ study with synthetic data).}
	\label{fig:F}
\end{figure}

To conclude the analysis of the models implemented with Fourier bases, Figure \ref{fig:F} presents the theoretical curves estimated by the best models according to the study metric.

In this $2^{\text{nd}}$ study with synthetic data, the results with Fourier bases are even better than those obtained with B-splines bases, showing the greater suitability of these bases when the curves present periodicity, as expected.

\subsection{Simulations}
\label{secao_replicacoes}

A study was carried out with replications to completely understand the behavior of the proposed method, in which we used the best model configurations according to the metric criterion \eqref{eq:r2} obtained in the studies conducted in Sections 3.3.1 and 3.3.2. Table \ref{tab:config} presents each of these configurations for better visualization of what they are according to the studies with synthetic data, error variance and basis functions.

\begin{table}[!htb]
	{\centering\fontsize{6}{17}\selectfont 
	\begin{tabular}{c|c|c|c}\hline
Configuration  & Data generated according to & $\sigma$ & Model fit specifications \\ \hline
1 & \multirow{2}{*}{$1^{\text{st}}$ study with synthetic data} & 0.1 & B-splines, $K=10$, $\mu=0.1$  \\
2 &  & 0.5 & B-splines, $K=10$, $\mu=0.1$ \\
\hline
3 & \multirow{4}{*}{$2^{\text{nd}}$ study with synthetic data} & 0.1 & B-splines, $K=15$, $\mu=0.3$  \\
4  && 0.5 & B-splines, $K=10$, $\mu=0.9$\\
5 & & 0.1 & Fourier bases, $K=30$, $\mu=0.04$ \\ 
6  &  & 0.5 & Fourier bases, $K=30$, $\mu=0.01$\\
\hline
	\end{tabular}
	\caption{Study configurations based on the best settings obtained in the $1^{\text{st}}$ and  $2^{\text{nd}}$ studies with synthetic data (Sections 3.3.1 and 3.3.2, respectively) according to the metric \eqref{eq:r2}.}
	\label{tab:config}
}
\end{table}


A total of 100 replications were performed for each configuration in Table \ref{tab:config}. We also considered $m=5$ for this study, so that the final average estimate of the coefficients ($\hat{\xi}_{k}=\frac{\vec{\hat{Z}_{k.}}'\vec{\hat{\beta}_{k.}}}{m}$, for the $k$-th coefficient) in each replication is given by the mean of the five final individual estimates. As the observed data are different in each replication and as there will be a single final average estimate of the coefficients per replication, it can be said that 100 final average estimates are obtained at the end of 100 replications for a given configuration in Table \ref{tab:config}, so that it is possible to estimate the distributions of the proposed metric \eqref{eq:r2} and the MSE, as well as the construction of boxplots for the final average estimates of the coefficients.

\begin{figure}[!htb]
	\centering
	\subfloat[Data with  $\sigma=0.1$\label{box1}]{\includegraphics[scale=0.32]{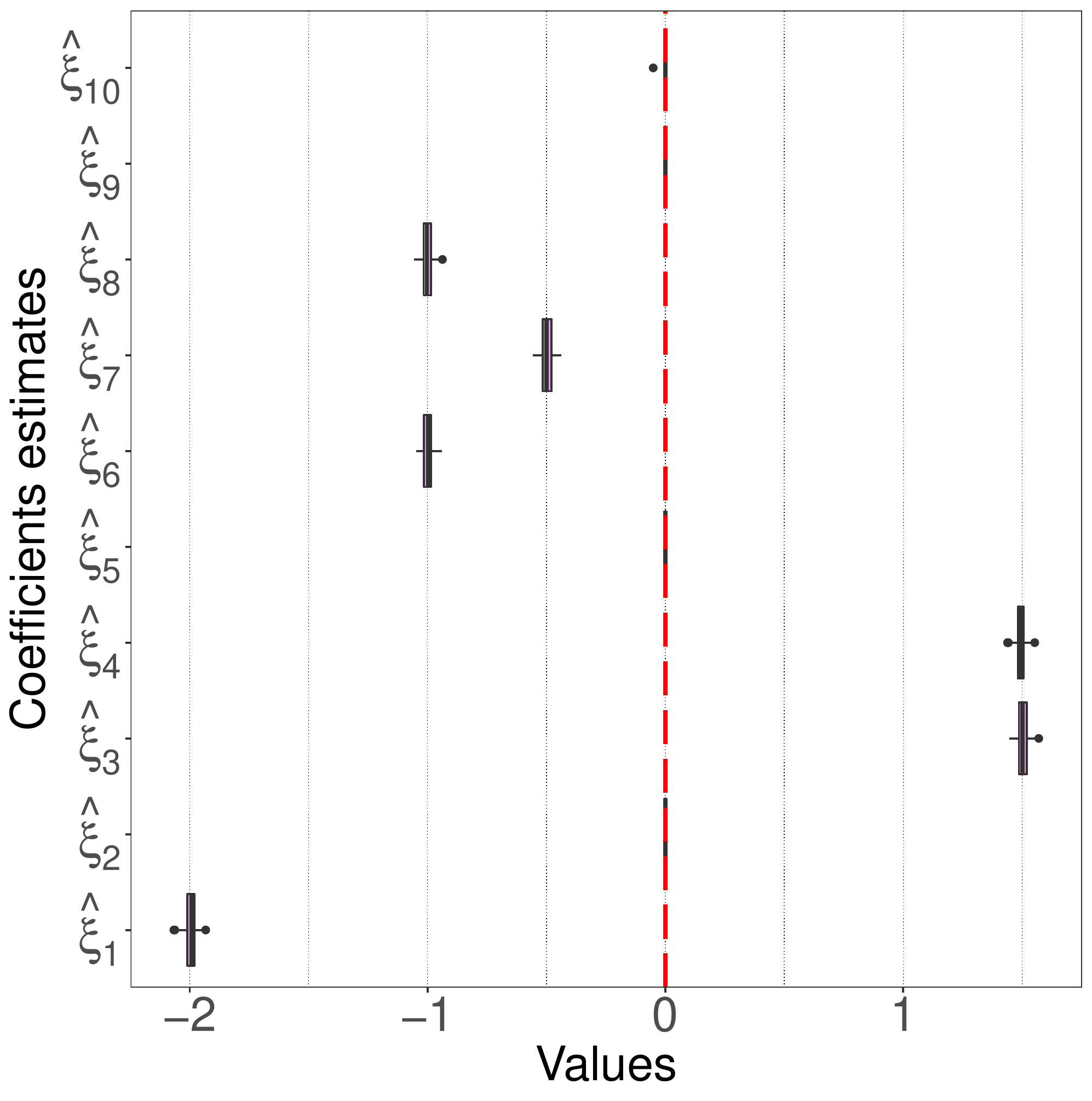}}
	\subfloat[Data with  $\sigma=0.5$\label{box2}]{\includegraphics[scale=0.32]{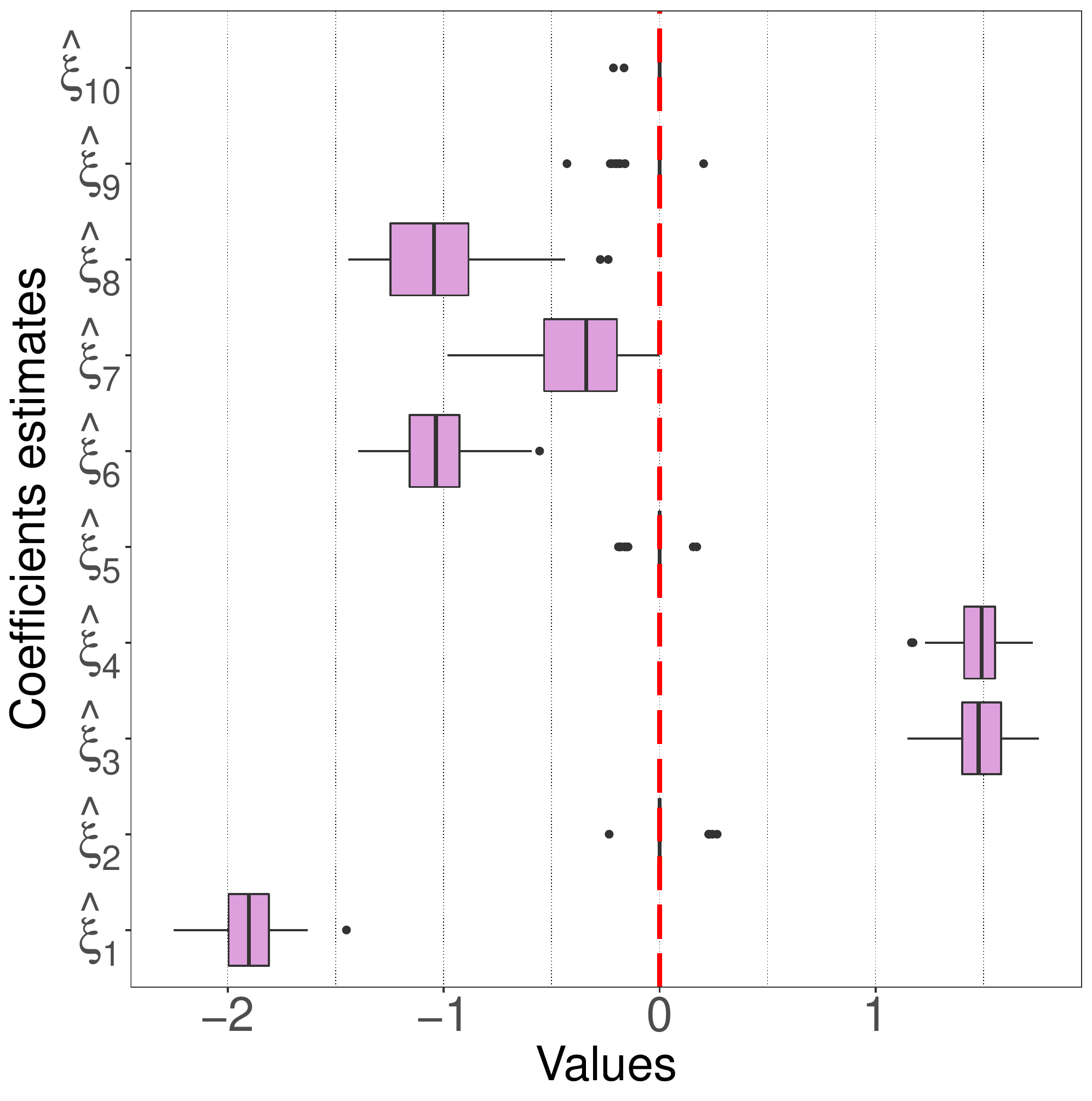}}
	\caption{\textit{Simulation configurations 1 and 2 (synthetic data by \eqref{1função} and B-spline bases for model fit). Boxplots of the final average coefficient estimates ($m=5$ curves) generated by the 100 replications, according to the data dispersion degree.}}
	\label{fig:boxplotc2}
\end{figure}

Therefore, Figure \ref{fig:boxplotc2} shows the boxplots of the final average estimates ($\hat{\xi}_{k}$'s) obtained for configurations 1 and 2 of Table \ref{tab:config}. It is easy to identify the coefficients that must be selected, since their boxplots are far away from zero, informing a high degree of significance regarding the non-nullity of these coefficients.

Due to the large amount of zeros in the replication results, the boxplots associated with the estimates $\hat{\xi}_{2}$, $\hat{\xi}_{5}$, $\hat{\xi}_{9}$ and $\hat{\xi}_{10}$ become practically invisible in Figure \ref{fig:boxplotc2}, and it is only possible to visualize the discrepant data points associated with these components, which are exactly those that were excluded from the linear combination that generated the data. It is also possible to notice that for both scenarios (data with $\sigma=0.1$ and with $\sigma=0.5$), there is a consensus on which basis functions should be excluded from the model.

\begin{figure}[!htb]
	\centering
	\subfloat[MSE’s]{\includegraphics[scale=0.32]{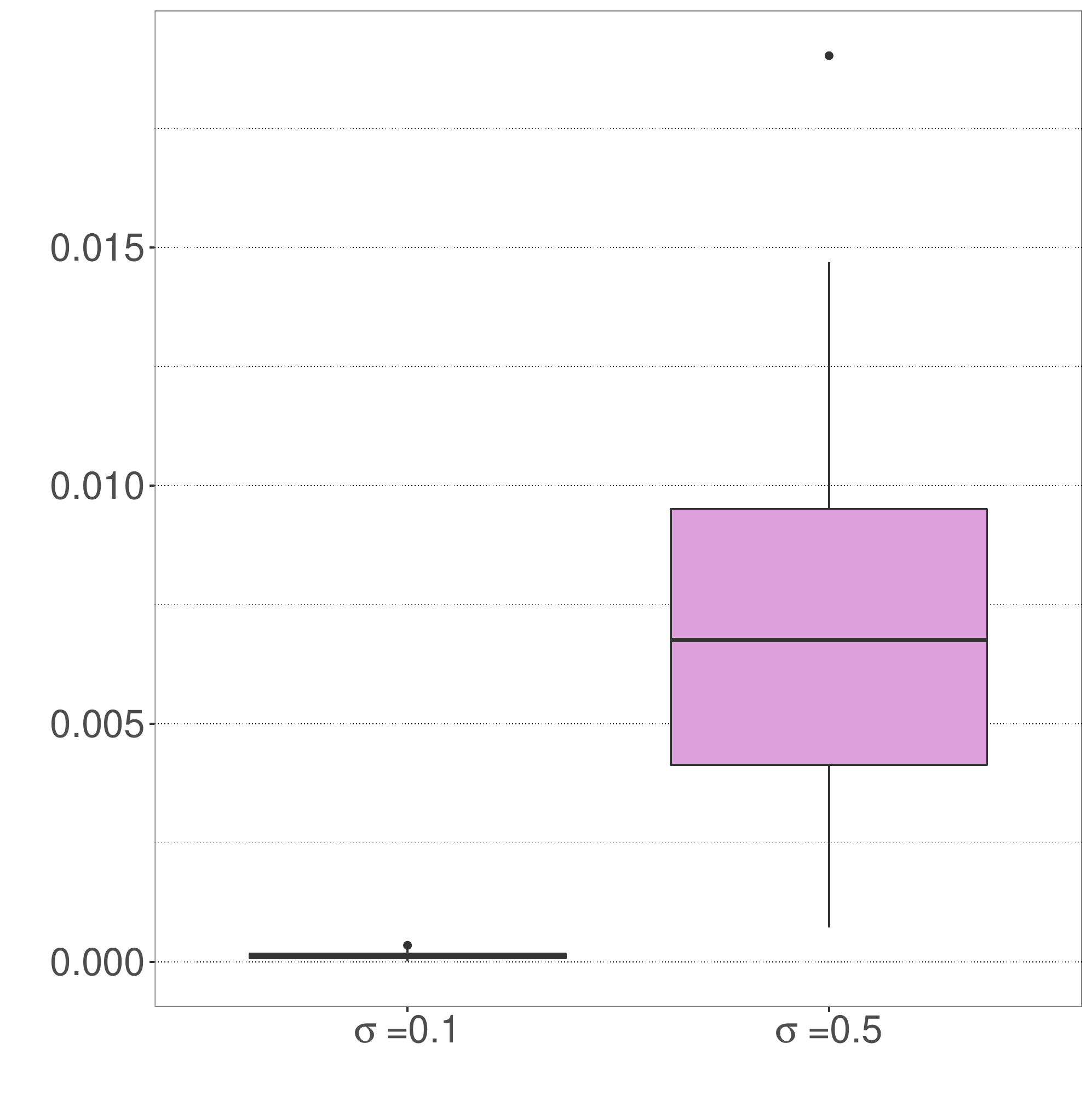}}
	\subfloat[Metric  \eqref{eq:r2}]{\includegraphics[scale=0.32]{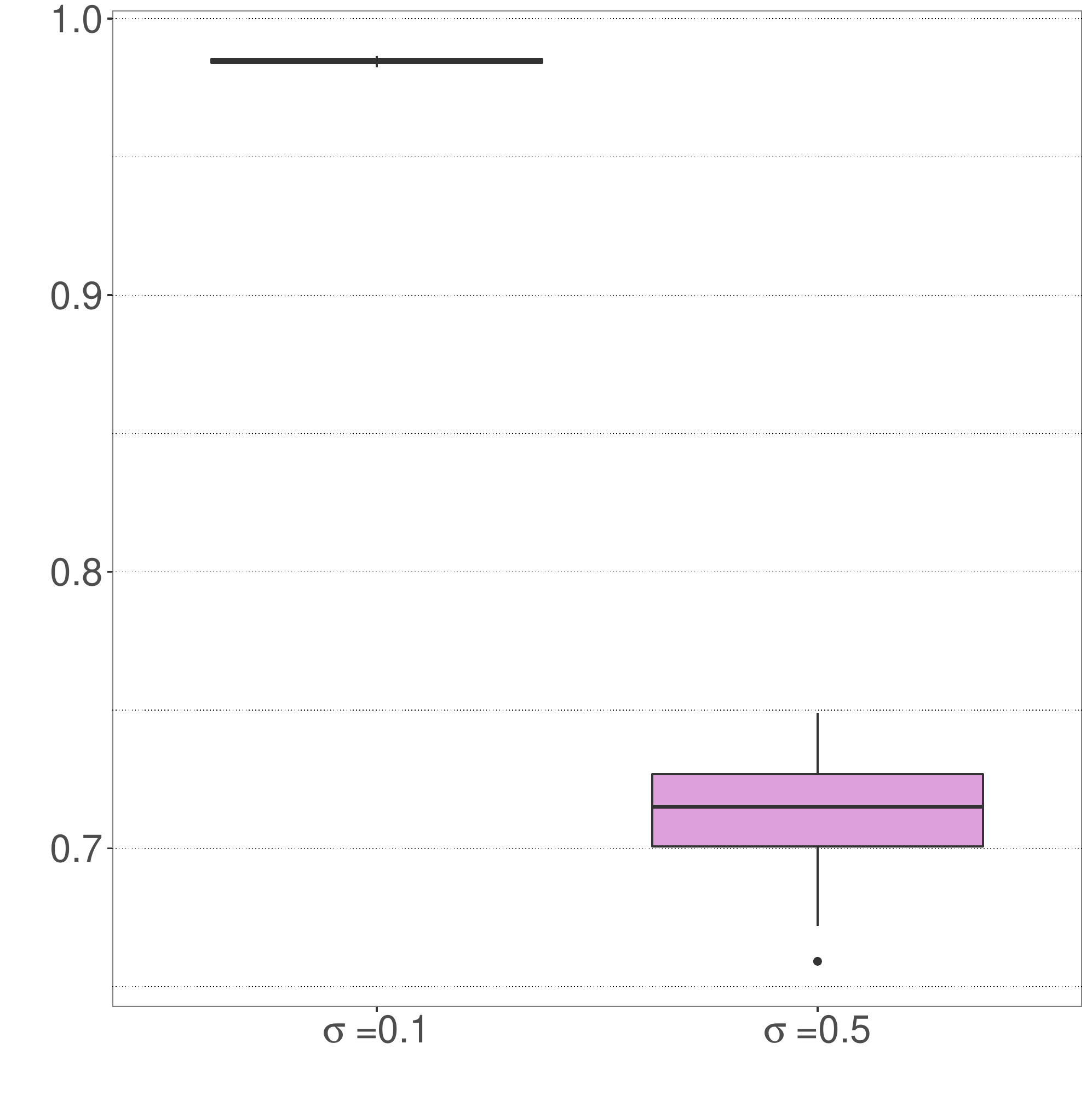}}
	\caption{\textit{Simulation configurations 1 and 2 (synthetic data by \eqref{1função} and B-spline bases for model fit)}. Boxplots of the performance metrics used according to the data dispersion degree.}
	\label{fig:metrics}
\end{figure}

Figure \ref{fig:metrics} shows the boxplots of the performance metrics for simulation configurations 1 and 2. Both metrics present expected behavior in relation to $\sigma$, since the performance of the models is better when the data presents low dispersion. It is also worth noting that both metrics are well behaved as they present few discrepant points and suggest symmetry.

\begin{figure}[!htb]
	\centering
	\subfloat[Data with  $\sigma=0.1$\label{b1}]{\includegraphics[scale=0.32]{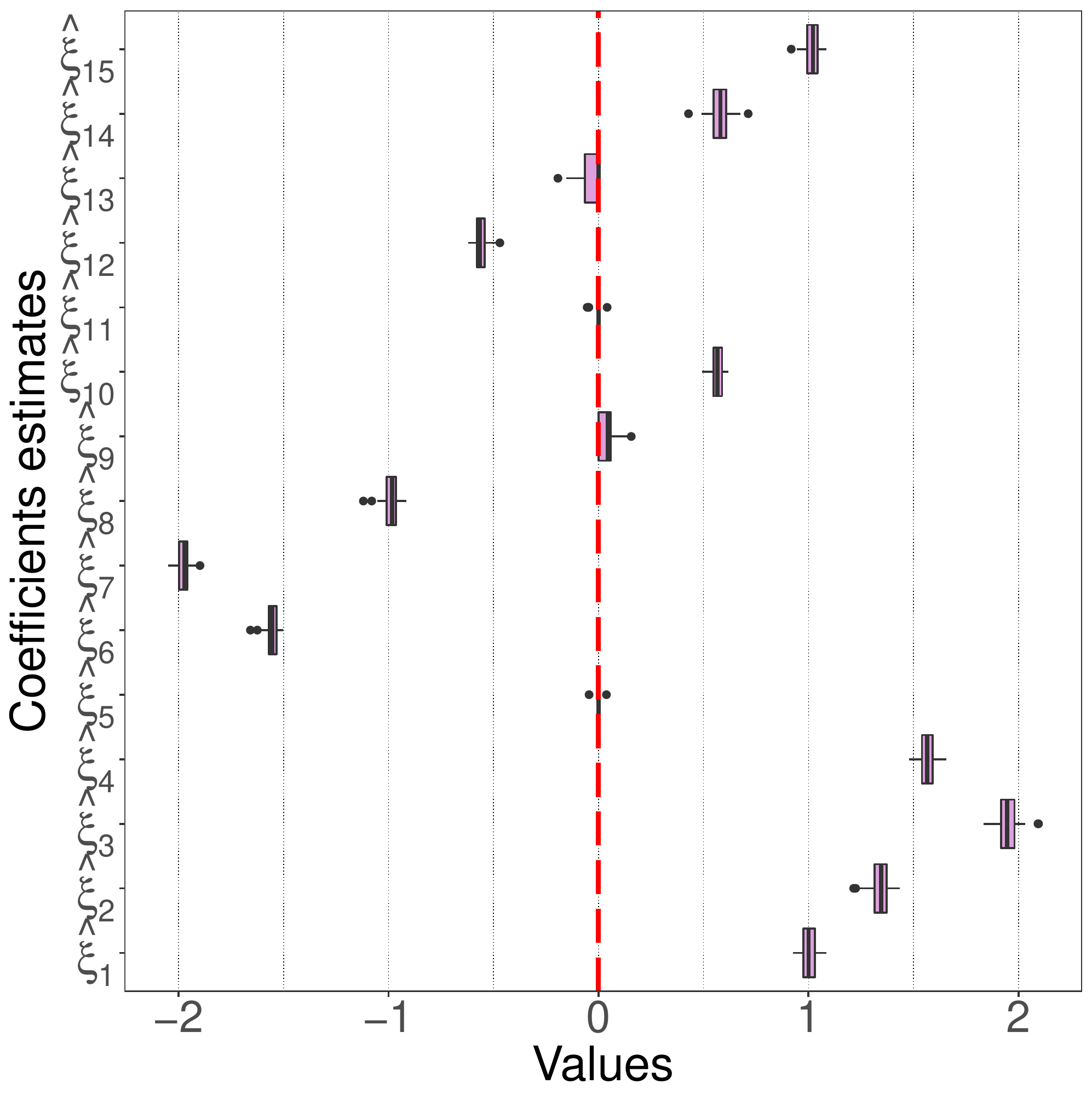}}
	\subfloat[Data with  $\sigma=0.5$\label{b2}]{\includegraphics[scale=0.32]{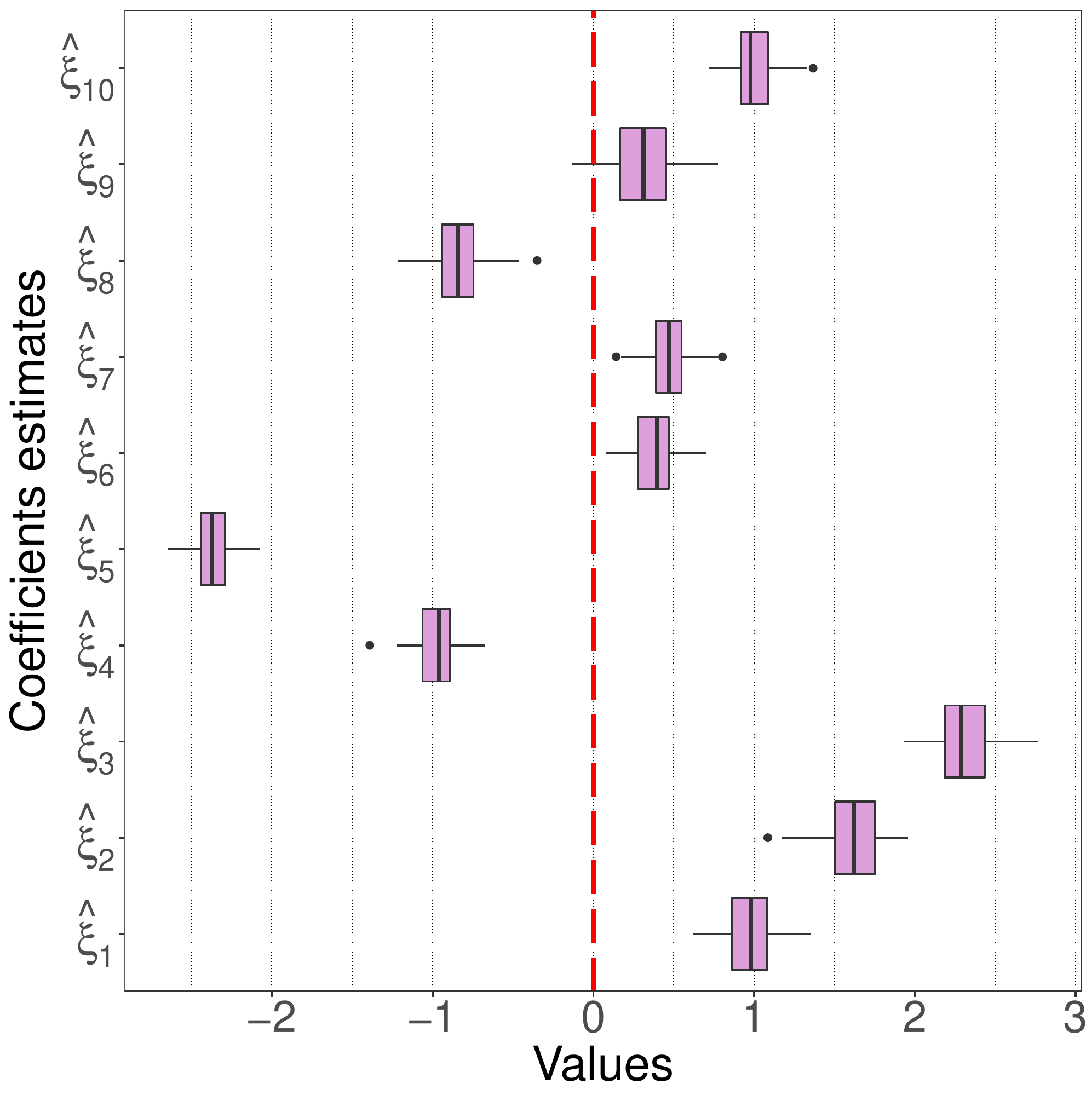}}
	\caption{\textit{Simulation configurations 3 and 4 (synthetic data by \eqref{eq:2} and B-spline bases for model fit)}. Boxplots of the final average coefficient estimates ($m=5$ curves) generated by the 100 replications, according to the data dispersion degree.}
	\label{fig:boxplotc1}
\end{figure}

Figure \ref{fig:boxplotc1} shows the results obtained by simulation configurations 3 and 4 (data generated according to $2^{\text{nd}}$ study with synthetic data and models fitted using B-splines). We can observe that the selected coefficients are associated with boxplots that have a higher interquartile deviation.

Unlike the results presented in Figure \ref{fig:boxplotc2}, in Figure \ref{fig:boxplotc1} it is not possible to carry out a comparative analysis between the scenario with $\sigma=0.1$ and the one with $\sigma=0.5$, since the $K$ parameter is different and the B-spline bases are not orthogonal. However, it is possible to see that there is no indication of exclusion of any basis in the scenario with $\sigma=0.5$, while although the model for $\sigma=0.1$ is initially defined with a larger number of bases, there is a clear indication that only 13  out of 15 bases should be selected. 

\begin{figure}[!htb]
	\centering
	\subfloat[MSE’s]{\includegraphics[scale=0.32]{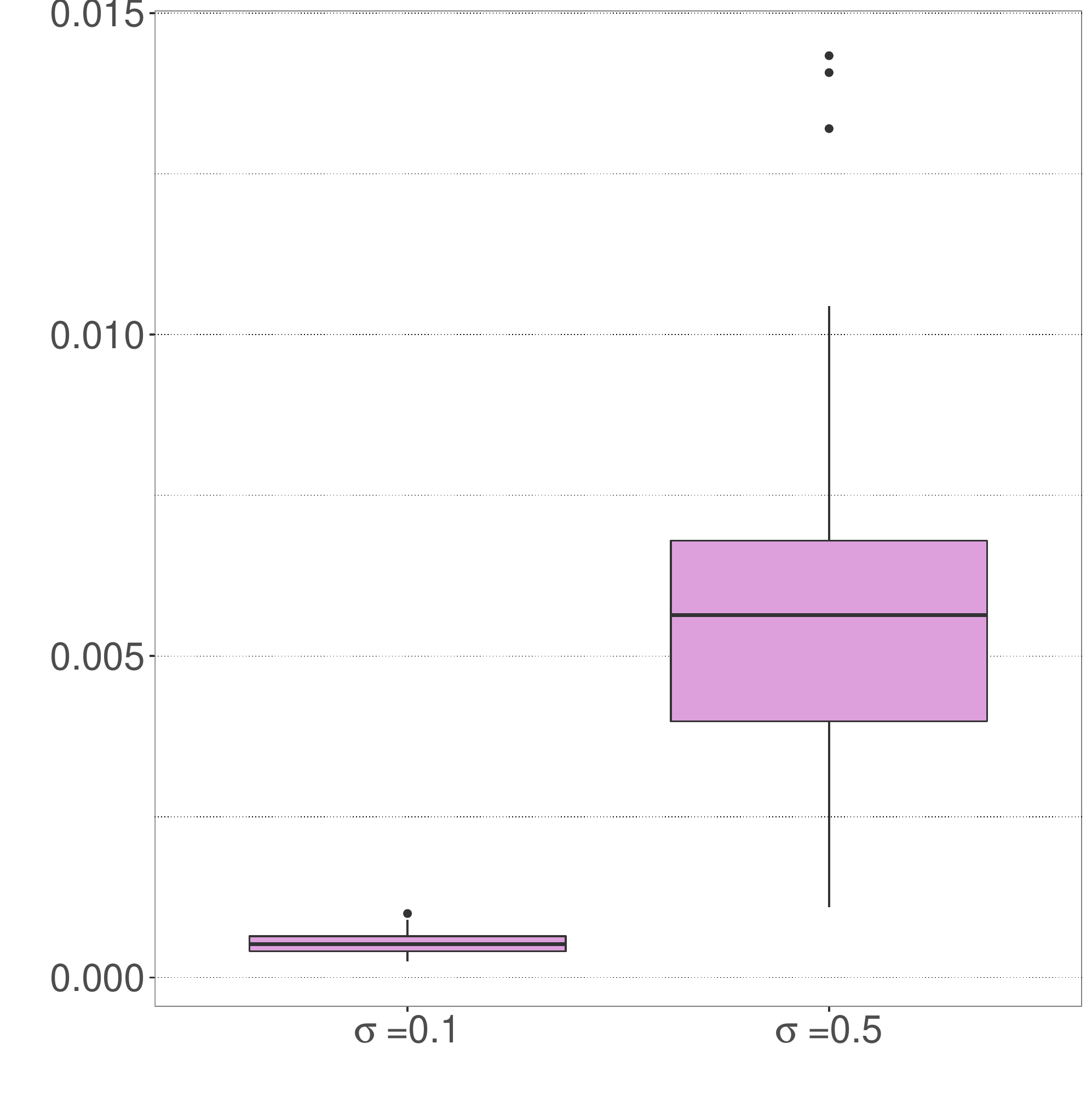}}
	\subfloat[Metric  \eqref{eq:r2}]{\includegraphics[scale=0.32]{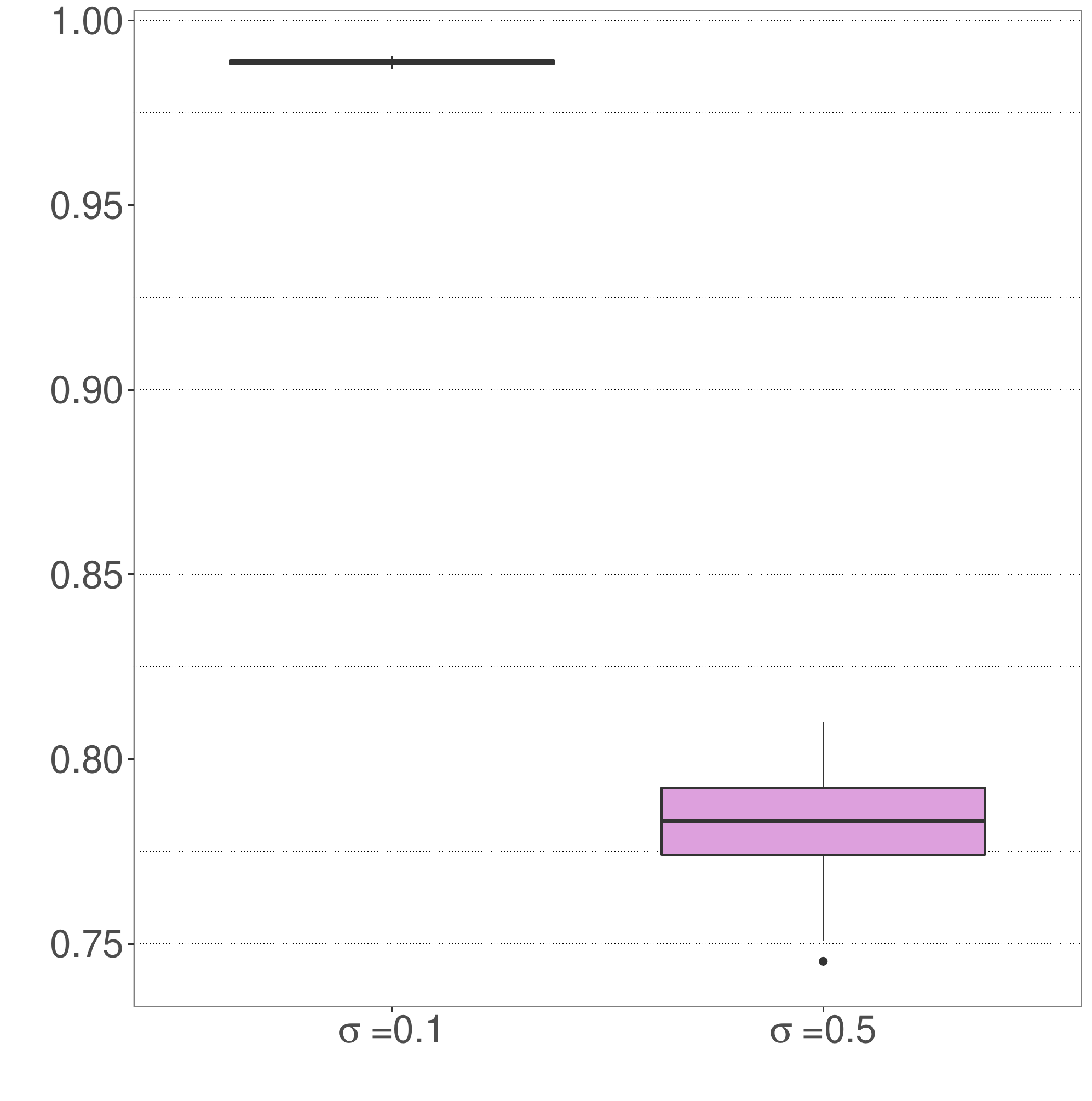}}
	\caption{\textit{Simulation configurations 3 and 4 (synthetic data by \eqref{eq:2} and B-spline bases for model fit)}. Boxplots of the performance metrics used according to the data dispersion degree.}
	\label{fig:metrics2}
\end{figure}

To conclude the analysis of the results of simulation configurations 3 and 4, Figure \ref{fig:metrics2} presents the boxplots of the performance metrics, showing again a better performance for $\sigma=0.1$ than for $\sigma=0.5$.

\begin{figure}[!htb]
	\centering
	\subfloat[Data with  $\sigma=0.1$]{\includegraphics[scale=0.34]{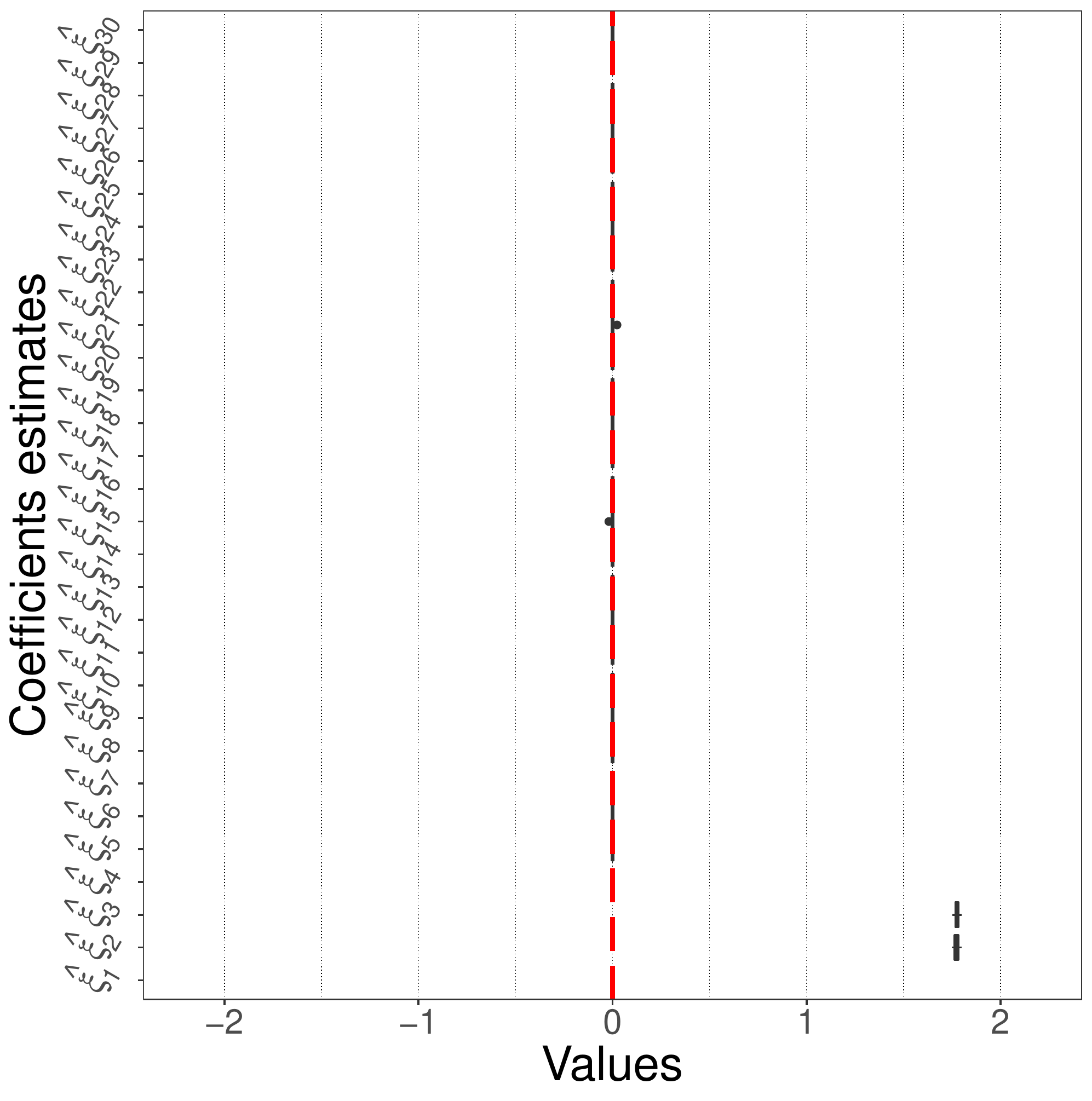}}
	\subfloat[Data with  $\sigma=0.5$]{\includegraphics[scale=0.34]{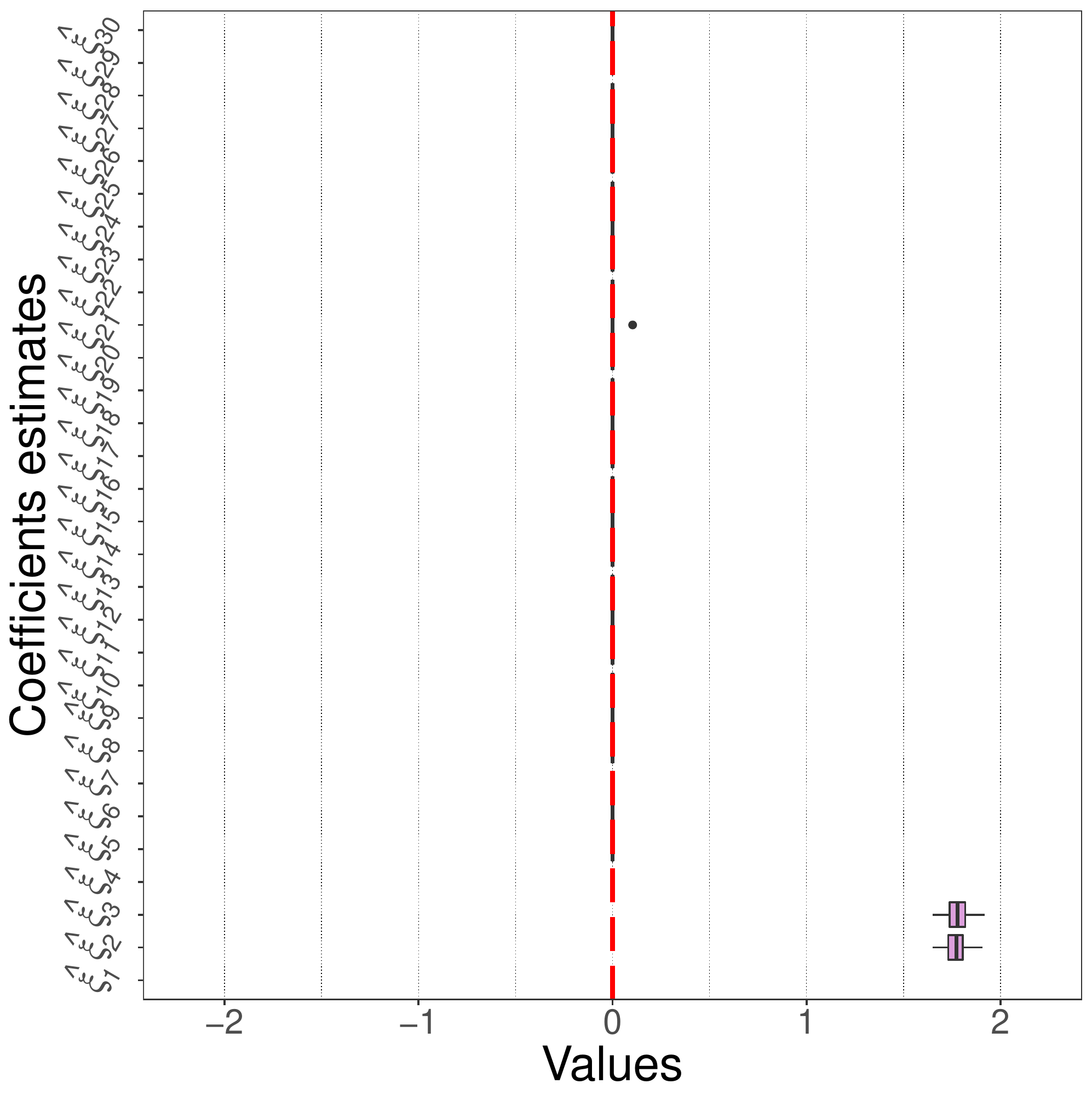}}
	\caption{\textit{Simulation configurations 5 and 6 (synthetic data by \eqref{eq:2} and Fourier bases for model fit)}. Boxplots of the final average coefficient estimates ($m=5$ curves) generated by the 100 replications, according to the data dispersion degree.}
	\label{fig:boxplotc1f}
\end{figure}

\begin{figure}[!htb]
	\centering
	\subfloat[MSE’s]{\includegraphics[scale=0.34]{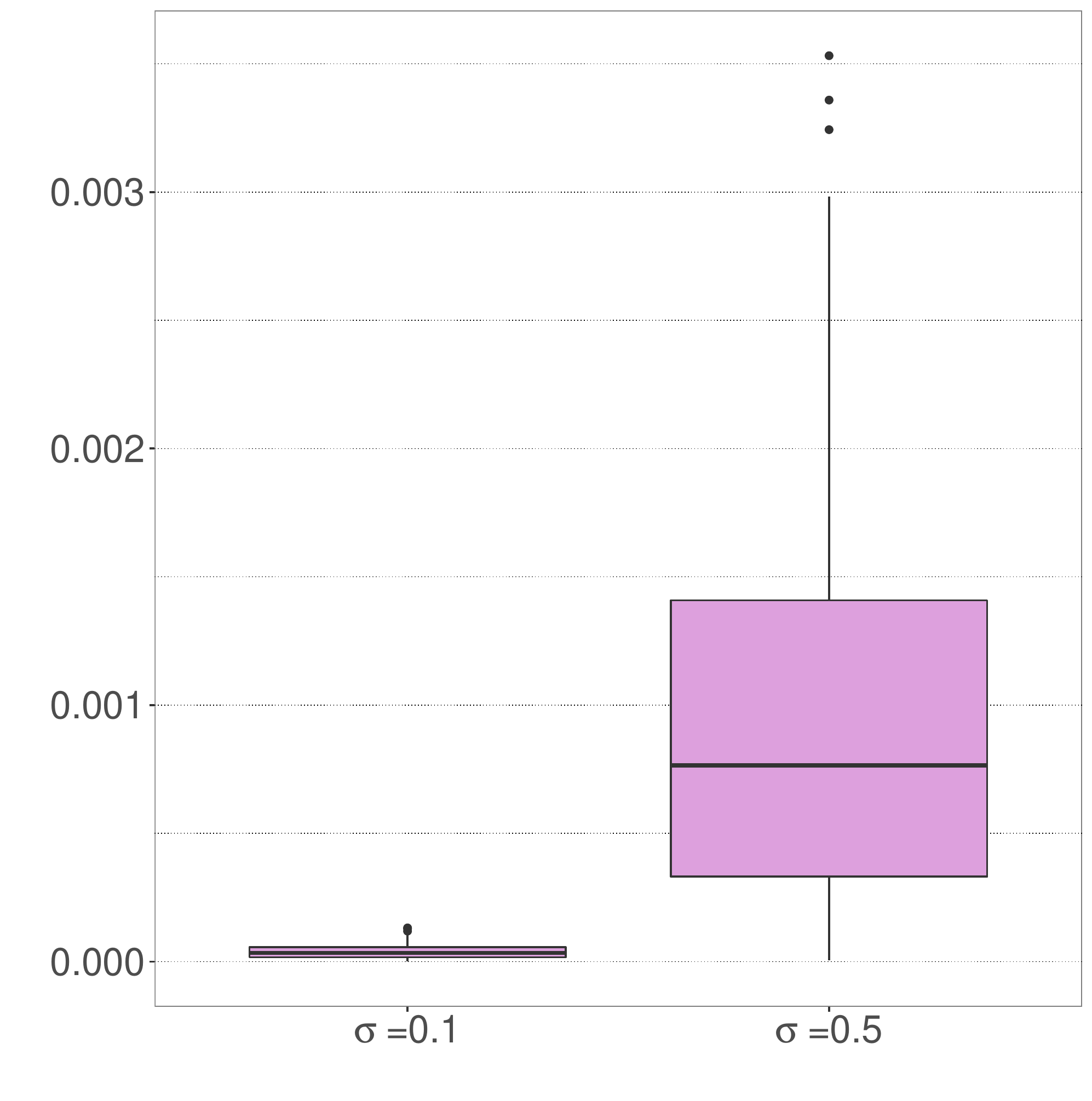}}
	\subfloat[Metric  \eqref{eq:r2}]{\includegraphics[scale=0.34]{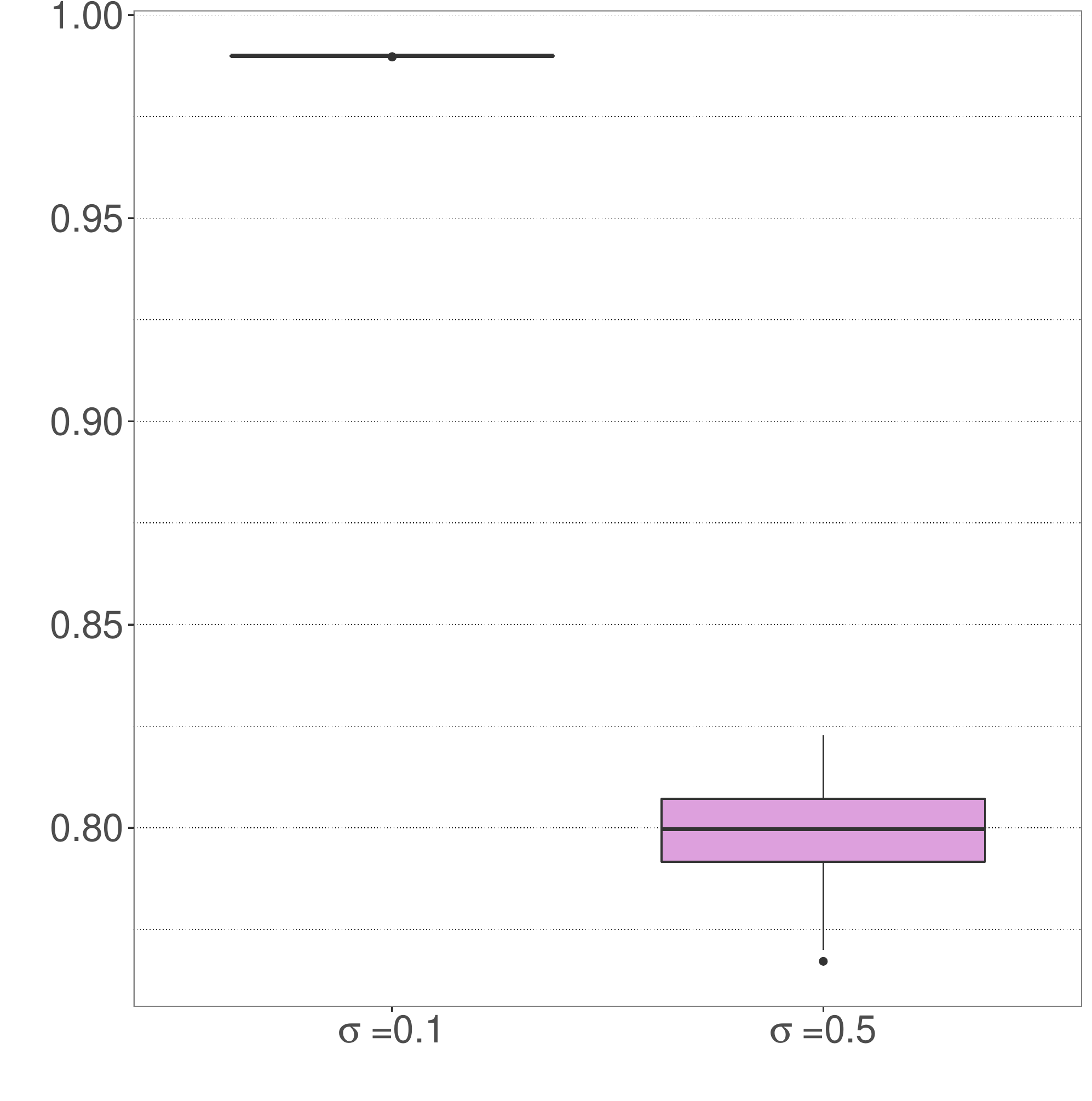}}
	\caption{\textit{Simulation configurations 5 and 6 (synthetic data by \eqref{eq:2} and Fourier bases for model fit)}. Boxplots of the performance metrics used according to the data dispersion degree.}
	\label{fig:metrics3}
\end{figure}

Finally, Figures \ref{fig:boxplotc1f} and \ref{fig:metrics3} present the results for simulation configurations 5 and 6. These results confirm the model’s ability to select two basis functions among many others, but also to obtain better fits when compared to models from simulation configurations 3 and 4. 


\subsection{LASSO and Bayesian Lasso versus the proposed model}

In addition to the smoothing and adaptive base selection method proposed in this paper, there are other methods that are also capable of inducing regularization and selection.

There are two methods in the statistical literature of linear regression that are widely known and that have the properties highlighted here: the LASSO frequentist method (Least Absolute Shrinkage and Selection Operator) presented by \citeauthor{lasso} in \citeyear{lasso}, and its Bayesian version, the Bayesian LASSO proposed by \citeauthor{art:bl-Casella} in \citeyear{art:bl-Casella}. Therefore, such techniques can be applied to functional data, by simply using the elements $B_{k}(\cdot)$'s where covariates are normally used. 

A brief comparative analysis between the proposed model and the two competing methods previously mentioned is presented in detail in the Supplementary Material considering $m=1$. The results show that, regardless of the metric being used (metric \eqref{eq:r2} or MSE), the superiority of the proposed model with $\mu$ as a parameter compared to Bayesian LASSO is remarkable. Compared with Bayesian LASSO, this version of the proposed model has enough of a satisfactory performance even to compete with LASSO, although a fairer comparison is restricted to evaluating LASSO against the proposed model with $\mu$ as a hyperparameter. In the latter case, it is also possible to note that both models have performances situated at the same level.

In analyzing the Supplementary Figures 3 to 12, it is possible to conclude that Bayesian LASSO has a lower selection power when compared to other competing models. In contrast, the proposed model accurately selects the base coefficients that were used in generating the synthetic data, both in the version that competes with Bayesian LASSO, as well as in the version that competes with LASSO. The LASSO frequentist method can also accurately select the base coefficients, although it is not able to select them correctly with the same frequency in the study with replications as the proposed model.

%% file: real_data.tex
\section{Application to real data and the GCV criterion}
\label{realdatacap}

For the evaluation of the proposed model in a set of real data, we chose to work with COVID-19 data, which are publicly available at the Brasil.IO open data portal (\citeauthor{BRASILio}, \citeyear{BRASILio}). This portal collects and structures data on the COVID-19 pandemic in Brazil through the information bulletins of the State Health Departments. Among the various variables present, the most important are confirmed cases of the disease and deaths. Therefore, the variable chosen for our analysis provides information on confirmed cases, and although the database is extremely large and includes data from states and municipalities, only information from the confirmed cases at the state level were used for this study.

Before running the proposed model, a little pre-processing of the data was performed. As data are recorded daily and as there is the possibility of recording errors associated with the process of updating the daily data, the total number of cases per week was taken as a variable for the study to apply the proposed model. As the first cases of COVID-19 in each state appeared in different weeks, it was decided to analyze the data from the $20^{\text{th}}$ epidemiological week of the year 2020, the week from which new cases were already registered in all members of the federation (states and Federal District). Starting from this week, the data extends to the $32^{\text{nd}}$ epidemiological week of the year 2021, for a total of 66 weeks.

It is necessary to emphasize that after pre-processing the data, the proposed model was adjusted considering $m=27$, meaning that the 26 states and the Federal District were fitted simultaneously, even though each state has its own curve pattern.

In the case of studies with synthetic data presented earlier, it is clearly known that the process generating the curves is the same. Thus, the performance metric \eqref{eq:r2} can naturally be used. However, when the curves have different generating processes and consequently present a substantially different behavior over the study interval, the metric \eqref{eq:r2} cannot be used as it is constructed taking into account the vector with the final average estimates of the coefficients ($\vec{\hat{\xi}}$). Therefore, it is necessary to resort to the alternative version of this metric, in which the final individual estimates are applied directly to the formula and $K_{\text{end}}$ is equivalent to $\sum_{k=1}^{K}Z_{ki}$. Therefore, considering $\vec{y_{i.}}=(y_{i1},\;y_{i2},\dots,\;y_{in_{i}})^{'}$, this alternative version of \eqref{eq:r2} can be calculated as the average of the quantities 

\begin{equation}
	1-\frac{(n_{i}-1)(\vec{y_{i.}}-\vec{B}\diag(\vec{\hat{Z}_{.i}})\vec{\hat{\beta}_{.i}})^{'}(\vec{y_{i.}}-\vec{B}\diag(\vec{\hat{Z}_{.i}})\vec{\hat{\beta}_{.i}})}{\left(n_{i}-\sum_{k=1}^{K}Z_{ki}\right)\left(\vec{y_{i.}}-\frac{1}{n_{i}}\sum_{j=1}^{n_{i}}y_{ij}\right)^{'}\left(\vec{y_{i.}}-\frac{1}{n_{i}}\sum_{j=1}^{n_{i}}y_{ij}\right)}\text{.}
	\label{eq:r2_alternativo_obs}
\end{equation}
\noindent across the curves, obtaining: 
\begin{equation}
  	1-\frac{1}{m}\sum_{i=1}^{m}	
  	\frac{(n_{i}-1)(\vec{y_{i.}}-\vec{B}\diag(\vec{\hat{Z}_{.i}})\vec{\hat{\beta}_{.i}})^{'}(\vec{y_{i.}}-\vec{B}\diag(\vec{\hat{Z}_{.i}})\vec{\hat{\beta}_{.i}})}{\left(n_{i}-\sum_{k=1}^{K}Z_{ki}\right)\left(\vec{y_{i.}}-\frac{1}{n_{i}}\sum_{j=1}^{n_{i}}y_{ij}\right)^{'}\left(\vec{y_{i.}}-\frac{1}{n_{i}}\sum_{j=1}^{n_{i}}y_{ij}\right)}
  	\text{.}
  	\label{eq:r2_alternativo}
\end{equation}In which: $\vec{\hat{Z}_{.i}}=(\hat{Z}_{1i},\hat{Z}_{2i},\dots,\hat{Z}_{Ki})^{'}$] and $\vec{\hat{\beta}_{.i}}=(\hat{\beta}_{1i},\hat{\beta}_{2i},\dots,\hat{\beta}_{Ki})^{'}$] are the estimates of the respective parameters of the $i$th observed curve. In the case of COVID-19 data, it is known that the evolution of the disease in each state started at different times of the year in 2020 and that the evolution of the curve for new cases directly depends on the sanitary measures adopted in each one. Therefore, it is necessary to use the metric \eqref{eq:r2_alternativo} to assess the overall goodness-of-fit of the model.

Before applying the proposed model, data on the number of new COVID-19 cases from each state had been scaled by dividing the $\vec{y_{i}}$s by their respective standard deviations. Thus, the estimated coefficients are associated with the data on this new scale. To recover the curves in their original scales, we need to multiply the curve estimate by the standard deviation of the observed data for the member of the Brazilian federation in question.

Our proposed Bayesian approach was implemented through the Gibbs sampler using two chains. The initialization of the chains, the maximum number of iterations, the definition of the \textit{burn-in} period, the spacing of points between the sampled values, as well as the definition of the hyperparameters and Bayesian estimates were all established following what is described in Section \ref{cap3seção1}. Furthermore, the diagnostic analysis based on the method proposed by \citeauthor{Gelman} (\citeyear{Gelman}) attested to the convergence of the partial coefficient' chains after the burn-in period in all tested model configurations.

Table \ref{tab:r2_alternativo_obs} presents the metric \eqref{eq:r2_alternativo_obs} values  obtained from applying the proposed model with $\mu=0.9$ and with B-spline bases on the COVID-19 data for each state in Brazil. The model fits the data well for most Brazilian states, with the best metric values across $K$ above 0.86. Among the 26 states of the federation and the Federal District, only six returned metric results between 0.64 and 0.86 for the best $K$. Only one state, Rio Grande do Norte (RN), resulted in a low metric value of 0.33051 ($K=30$).

{\centering\fontsize{6}{17}\selectfont 
	\begin{longtable}[c]{c|ccccc}
	\caption{Values of metric \eqref{eq:r2_alternativo_obs} for each member of the federation, according to the number $K$ of B-spline bases used.}
	\label{tab:r2_alternativo_obs}\endfirsthead\endhead\hline
		\multicolumn{1}{c}{\multirow{2}{*}{Member}} & \multicolumn{5}{c}{$K$}                           \\\cline{2-6}
		\multicolumn{1}{c}{}                        & 10      & 15      & 20      & 25      & 30      \\\hline
		AC                                          & 0.79727 & 0.85847 & 0.89068 & 0.89786 & 0.88261 \\
		AL                                          & 0.84784 & 0.87407 & 0.92056 & 0.92493 & 0.92660 \\
		AM                                          & 0.76053 & 0.84699 & 0.89943 & 0.95554 & 0.95353 \\
		AP                                          & 0.58033 & 0.63509 & 0.69188 & 0.72406 & 0.74154 \\
		BA                                          & 0.78685 & 0.81881 & 0.81083 & 0.86920 & 0.87727 \\
		CE                                          & 0.90476 & 0.90717 & 0.90246 & 0.89743 & 0.89079 \\
		DF                                          & 0.86535 & 0.91255 & 0.91636 & 0.94577 & 0.95094 \\
		ES                                          & 0.74622 & 0.86217 & 0.90367 & 0.91064 & 0.91474 \\
		GO                                          & 0.80756 & 0.82012 & 0.83604 & 0.88103 & 0.87018 \\
		MA                                          & 0.88484 & 0.93278 & 0.91934 & 0.92536 & 0.92452 \\
		MG                                          & 0.86819 & 0.89613 & 0.95330 & 0.96150 & 0.95470 \\
		MS                                          & 0.75386 & 0.90799 & 0.95633 & 0.95700 & 0.94682 \\
		MT                                          & 0.84478 & 0.84733 & 0.89996 & 0.91444 & 0.89697 \\
		PA                                          & 0.83569 & 0.84393 & 0.85816 & 0.87363 & 0.86947 \\
		PB                                          & 0.73202 & 0.85386 & 0.93241 & 0.91833 & 0.93652 \\
		PE                                          & 0.89442 & 0.90629 & 0.92193 & 0.92226 & 0.91310 \\
		PI                                          & 0.91887 & 0.92230 & 0.94238 & 0.95768 & 0.95657 \\
		PR                                          & 0.51133 & 0.64714 & 0.65424 & 0.67069 & 0.74477 \\
		RJ                                          & 0.53561 & 0.65148 & 0.64051 & 0.64450 & 0.67990 \\
		RN                                          & 0.20451 & 0.22261 & 0.16066 & 0.14595 & 0.33051 \\
		RO                                          & 0.81777 & 0.82672 & 0.87787 & 0.90532 & 0.92070 \\
		RR                                          & 0.55322 & 0.65284 & 0.61558 & 0.64986 & 0.59560 \\
		RS                                          & 0.53839 & 0.63156 & 0.58727 & 0.68203 & 0.70597 \\
		SC                                          & 0.54724 & 0.73610 & 0.71856 & 0.71667 & 0.71161 \\
		SE                                          & 0.76553 & 0.86625 & 0.87400 & 0.87727 & 0.89405 \\
		SP                                          & 0.81712 & 0.85062 & 0.91171 & 0.92095 & 0.91873 \\
		TO                                          & 0.69393 & 0.87561 & 0.90200 & 0.94343 & 0.93819\\\hline
	\end{longtable}
}

Table \ref{tab:r2_alternativo_geral} shows the global results for each scenario of $K$, taking into account the average among the states, which is characterized by the metric \eqref{eq:r2_alternativo}.

\begin{table}[!htb]
\centering
	\begin{tabular}{c|ccccc}\hline
	 $K$&10      & 15      & 20      & 25      & 30      \\\hline
	Metric  \eqref{eq:r2_alternativo}&0.73385&	0.80026&	0.81845&	0.83679&	0.84618
	  	\\\hline
	\end{tabular}
	\caption{Metric \eqref{eq:r2_alternativo} values for each number of basis functions $K$ considered.}
\label{tab:r2_alternativo_geral}
\end{table}

As much as the metric in question is built based on the penalty of excess parameters, it is clear that the goodness-of-fit according to it increases for the highest values of $K$ tested. However, by varying $K$ by 5 units, it is noted that the gain in terms of explained variability decreases as the $K$ value increases. In other words, the best model is found between $K=25$ and $K=30$, since the metric will hardly present significantly larger values for larger $K$ values.

In the case of studies with synthetic data, the Mean Square Error (MSE) criterion becomes a good measure to assess the predictive capacity of the model since it is built from the distances between points of the theoretical curve that generated the data and estimated curve points. Unfortunately, this way of calculating the MSE is not possible when working with real data due to the lack of knowledge about the function that generated the data, leaving the possibility of calculating the MSE from the distances between the observed values and the estimated curve points. The major problem in calculating the MSE from the observed data lies in the fact that this measure is not able to identify models which are overfitted to the observed data (overfitting), and it is necessary to combine cross-validation techniques with the MSE to overcome this problem.

Although the conventional cross-validation associated with the MSE is an effective procedure to circumvent the overfitting problem, it often becomes inefficient due to the high computational cost. As an alternative, Generalized Cross-Validation (GCV) (\citeauthor{Wahba}, \citeyear{Wahba}) is defined, which is nothing more than a modified form of cross-validation based on minimizing the following objective function: 

\begin{equation}
	\GCV{}_{i}(\lambda)=\frac{1}{n}\frac{\sum_{j=1}^{n}\left[y_{ij}-\hat{x}(t_{ij})\right]^2}{[1-\frac{1}{n}\tr(\vec{S_{\lambda}})]^2}\text{.}
	\label{VCG}
\end{equation}

\noindent Adapting Expression \eqref{VCG} to the proposed model, we obtain:
\begin{equation}
	\GCV{}_{i}(K)=\frac{1}{n_{i}}\frac{(\vec{y_{i.}}-\vec{B}\diag(\vec{\hat{Z}_{.i}})\vec{\hat{\beta}_{.i}})^{'}(\vec{y_{i.}}-\vec{B}\diag(\vec{\hat{Z}_{.i}})\vec{\hat{\beta}_{.i}})}{[1-\frac{1}{n_{i}}\tr(\vec{S_{K}})]^2},
	\label{eq:VCG}
\end{equation}

\noindent where $\vec{S_{K}}$ is the projection matrix of the proposed model with $K$ candidate bases.

To understand how $\vec{S_{K}}$ is calculated, it is necessary to recall that the mean of the full conditional distribution of $\vec{\beta_{.i}}$, which is equivalent to the mode since the normal density is symmetric, is given by: 
\begin{equation}
		\E(\vec{\beta_{.i}}|\vec{\beta_{.-i}},\vec{\theta},\vec{\mu},\sigma^2,\tau^2,\vec{Z},\vec{Y})=\vec{D_{i}}^{-1}\vec{G_{i.}}^{'}\vec{y_{i.}}\text{,}\nonumber
\end{equation}as shown in Appendix \ref{A_full} Expression (\ref{eq:mean_beta_i}).

Finally, when considering $\vec{\hat{\beta}_{.i}}=\vec{\hat{D}_{i}}^{-1}\vec{\hat{G}_{i.}}^{'}\vec{y_{i.}}$, in which $\vec{\hat{D}_{i}}$ and $\vec{\hat{G}_{i.}}$ are the respective estimates of $\vec{D_{i}}$ and $\vec{G_{i.}}$, then the estimated curve is represented by: 
\begin{equation}
	\vec{\hat{y}_{i.}}=\vec{B}\diag(\vec{\hat{Z}_{.i}})\vec{\hat{\beta}_{.i}}=\vec{B}\diag(\vec{\hat{Z}_{.i}})\vec{\hat{D}_{i}}^{-1}\vec{\hat{G}_{i.}}^{'}\vec{y_{i.}}\text{,}\nonumber
\end{equation}but $\vec{B}\diag(\vec{\hat{Z}_{.i}})=\vec{\hat{G}_{i.}}$, so that the projection matrix can be represented by:
\begin{equation}
	\vec{S_{K}}=\vec{\hat{G}_{i.}}\vec{\hat{D}_{i}}^{-1}\vec{\hat{G}_{i.}}^{'}\nonumber
\end{equation}

\noindent Thus, it is completely possible to calculate the $\GCV{}_{i}(.)$ for all models tested as a function of $K$.

Table \ref{tab:VCG} shows the values of the Generalized Cross-Validation criterion obtained from implementing the proposed model with $\mu=0.9$ and with B-spline bases for each state.

\begin{table}[!htb]
	{\centering\fontsize{6}{17}\selectfont 
	\begin{tabular}{c|ccccc}\hline
				\multicolumn{1}{c|}{\multirow{2}{*}{Member}} & \multicolumn{5}{c}{K}                           \\\cline{2-6}
				\multicolumn{1}{c|}{}                        & 10      & 15      & 20      & 25      & 30      \\\hline
				AC                                          & 0.23612 & 0.17909 & 0.15099 & 0.15540 & 0.19846 \\
				AL                                          & 0.17722 & 0.15935 & 0.10972 & 0.11422 & 0.12409 \\
				AM                                          & 0.27891 & 0.19362 & 0.13891 & 0.06765 & 0.07856 \\
				AP                                          & 0.48878 & 0.46176 & 0.42558 & 0.41983 & 0.43695 \\
				BA                                          & 0.24825 & 0.22928 & 0.26129 & 0.19901 & 0.20748 \\
				CE                                          & 0.11092 & 0.11747 & 0.13472 & 0.15605 & 0.18463 \\
				DF                                          & 0.15682 & 0.11067 & 0.11553 & 0.08251 & 0.08293 \\
				ES                                          & 0.29557 & 0.17441 & 0.13305 & 0.13596 & 0.14414 \\
				GO                                          & 0.22413 & 0.22762 & 0.22646 & 0.18101 & 0.21947 \\
				MA                                          & 0.13412 & 0.08506 & 0.11140 & 0.11356 & 0.12761 \\
				MG                                          & 0.15351 & 0.13144 & 0.06450 & 0.05857 & 0.07658 \\
				MS                                          & 0.28667 & 0.11643 & 0.06032 & 0.06542 & 0.08991 \\
				MT                                          & 0.18078 & 0.19319 & 0.13817 & 0.13018 & 0.17419 \\
				PA                                          & 0.19136 & 0.19749 & 0.19591 & 0.19227 & 0.22067 \\
				PB                                          & 0.31211 & 0.18493 & 0.09336 & 0.12426 & 0.10732 \\
				PE                                          & 0.12297 & 0.11858 & 0.10783 & 0.11828 & 0.14692 \\
				PI                                          & 0.09449 & 0.09833 & 0.07959 & 0.06439 & 0.07342 \\
				PR                                          & 0.56914 & 0.44651 & 0.47758 & 0.50102 & 0.43148 \\
				RJ                                          & 0.54087 & 0.44101 & 0.49653 & 0.54088 & 0.54115 \\
				RN                                          & 0.92648 & 0.98372 & 1.15932 & 1.29940 & 1.13182 \\
				RO                                          & 0.21223 & 0.21927 & 0.16868 & 0.14406 & 0.13406 \\
				RR                                          & 0.52035 & 0.43930 & 0.53097 & 0.53273 & 0.68366 \\
				RS                                          & 0.53762 & 0.46623 & 0.57007 & 0.48377 & 0.49707 \\
				SC                                          & 0.52732 & 0.33394 & 0.38873 & 0.43107 & 0.48755 \\
				SE                                          & 0.27308 & 0.16925 & 0.17404 & 0.18674 & 0.17911 \\
				SP                                          & 0.21300 & 0.18902 & 0.12194 & 0.12026 & 0.13739 \\
				TO                                          & 0.35648 & 0.15740 & 0.13536 & 0.08606 & 0.10449\\\hline
	\end{tabular}
	
	\caption{Criterion values \eqref{eq:VCG} for each member of the federation according to the $K$ value used.}
\label{tab:VCG}
}
\end{table}

The criterion \eqref{eq:VCG} was constructed to choose the best model once the state is fixed, meaning the best number of B-spline bases ($K$) to be used in the proposed model. However, it is interesting to note that the comparison of $\GCV$s between states enables us to observe the same groups of results found in the conclusions that were obtained through the metric \eqref{eq:r2_alternativo_obs}, which is considered the best scenario for each state concerning $K$ among the 26 states of the federation and the Federal District. Thus, the same six that returned results between 0.64 and 0.86 for the metric \eqref{eq:r2_alternativo_obs} now return values between 0.2 and 0.47 for the generalized cross-validation criterion \eqref{eq:VCG}. Again, the state of Rio Grande do Norte (RN), whose data were not well-fitted by the proposed model, returned a $\GCV$ of 0.92648, the worst of all. The $\GCV$s for the other states are all at a level below 0.2.

As already highlighted, the main objective of the generalized cross-validation criterion \eqref{eq:VCG} is to point to the choice of the best $K$. This can be done separately for each state, but it can also be done in a more generalized way through the following expression: 

\begin{equation}
	\GCV(K)=\frac{1}{m}\sum_{i=1}^{m}\GCV{}_{i}(K)\text{.}	
	\label{eq:VCG2}
\end{equation}

Table \ref{tab:VCG2} shows the overall results for each $K$ scenario taking into account the average among the states, which is characterized by the criterion \eqref{eq:VCG2}. We can observe a gain in fit quality as $K$ increases for the first four columns of the table. However, the GCV criterion increases from $K=25$ onwards, suggesting that the best model is obtained when $K=25$.

\begin{table}[!htb]
	\centering
	\begin{tabular}{c|ccccc}\hline
		K&10      & 15      & 20      & 25      & 30      \\\hline
		Criterion \eqref{eq:VCG2}&0.30997&	0.25275&	0.25076&	0.24832&	0.26004
		\\\hline
	\end{tabular}
	\caption{Criterion values \eqref{eq:VCG2} according to the $K$ value used.}
	\label{tab:VCG2}
\end{table}

\begin{figure}[!htb]
	\centering
	\subfloat[São Paulo, SP]{\includegraphics[scale=0.25]{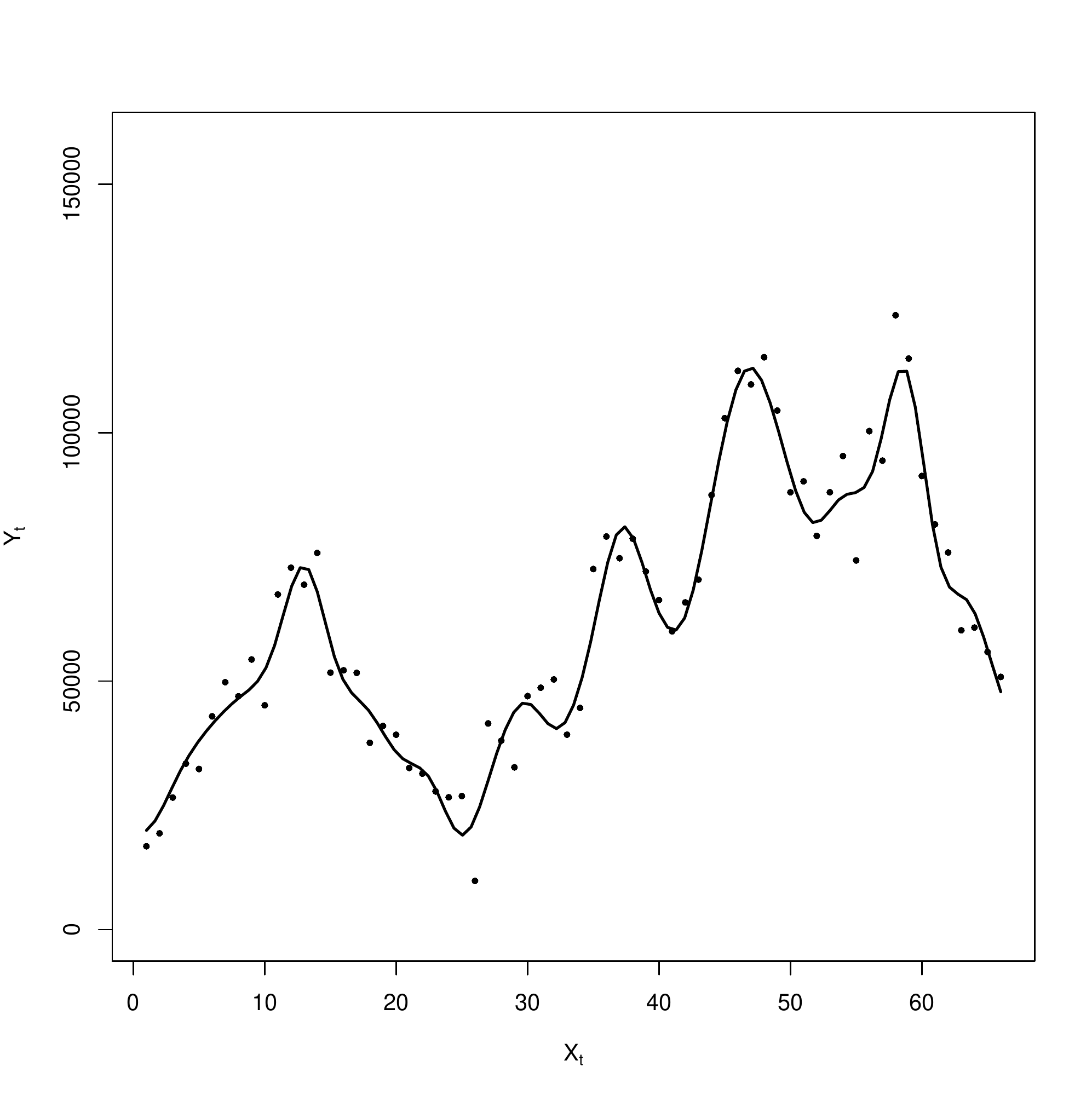}}
	\subfloat[Rio Grande do Sul, RS]{\includegraphics[scale=0.25]{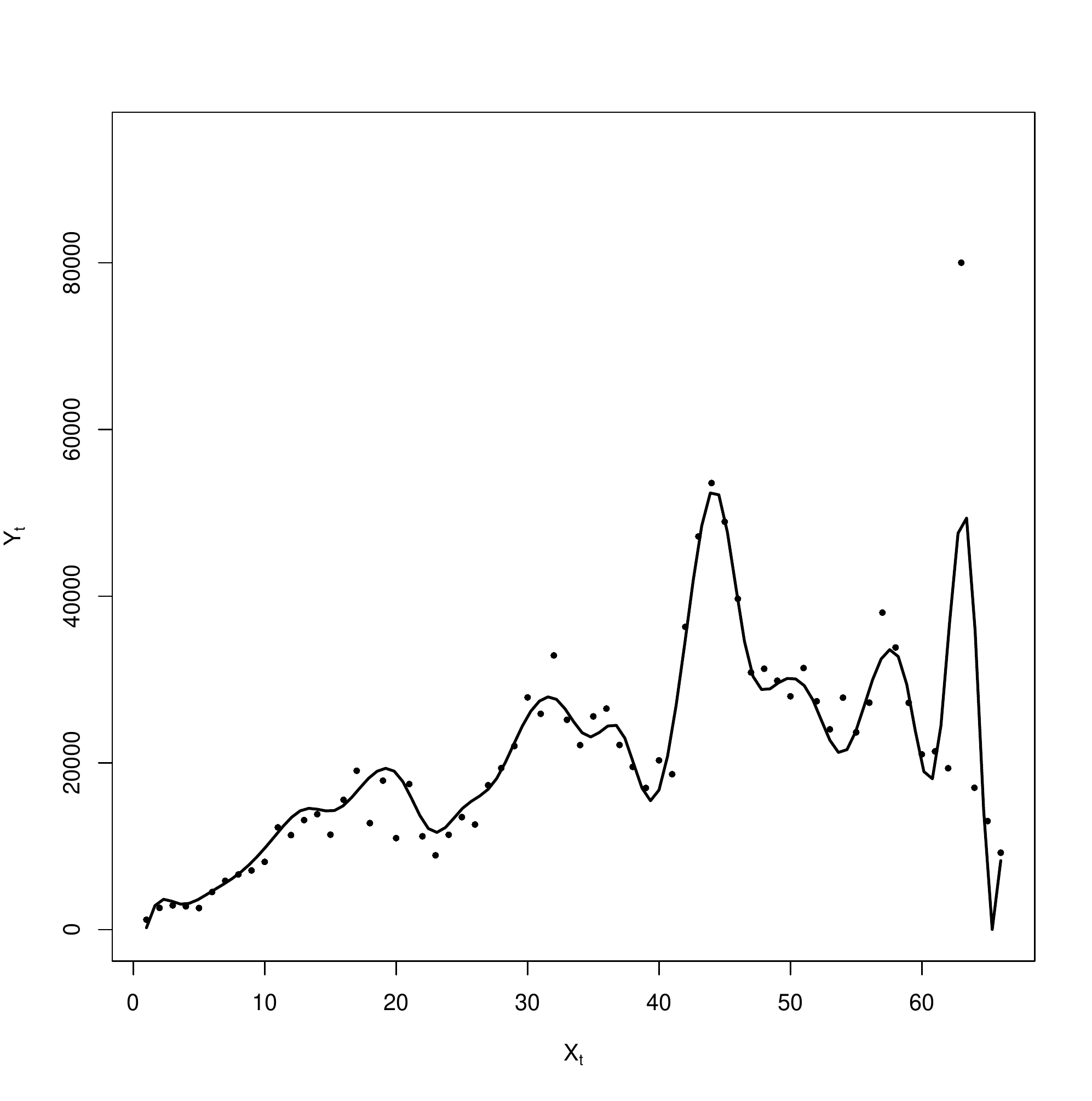}}
	\subfloat[Federal District, FD]{\includegraphics[scale=0.25]{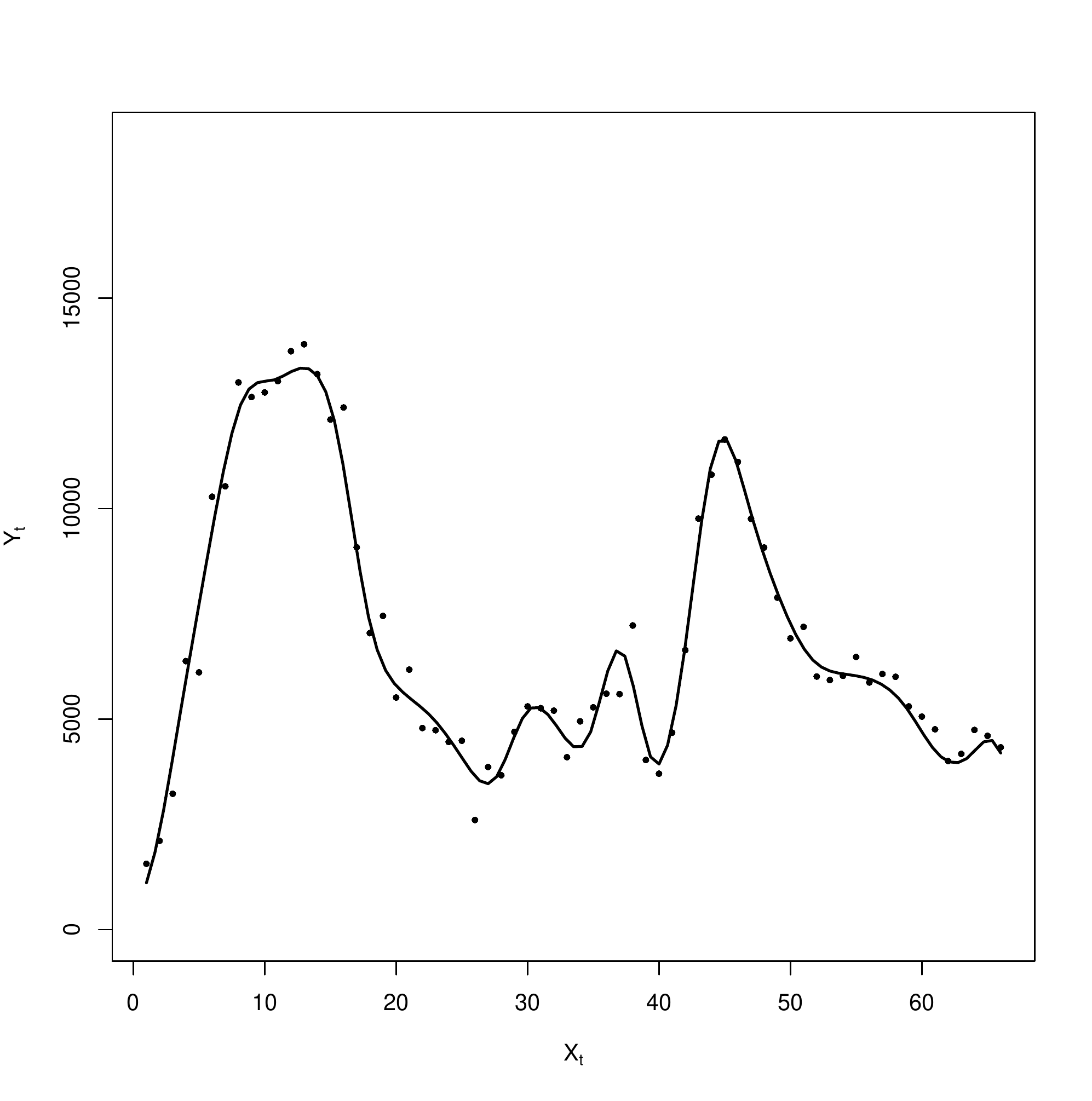}}
	\caption{Estimated curves of weekly new cases of COVID-19 for three members of the Brazilian federation (proposed model with $K=30$ B-splines).
	\label{3curvas}
}	
\end{figure}

As an example, Figure \ref{3curvas} shows the fitted curves for three members of the Brazilian federation in order to visually assess the model fit. The rescaled data were used for the calculation of both metric \eqref{eq:r2_alternativo_obs}  and GCV criterion \eqref{eq:VCG}. In addition, the data scale was retrieved to generate the graphs in Figure \ref{3curvas}. Data from the state of São Paulo (SP) and the Federal District (FD) were well fitted by the proposed model as already verified by the metric \eqref{eq:r2_alternativo_obs} in Table \ref{tab:r2_alternativo_obs}. However, Rio Grande do Sul (RS), which is in the group of 6 states with values below 0.86 for the metric \eqref{eq:r2_alternativo_obs}, shows an estimated curve that cannot fit well the observations corresponding to the last weeks in the dataset due to the existence of a highly influential outlier.

%% file: discussion.tex
\section{Conclusion }

An open problem in functional data analysis is determining which bases are preferable in a model among a set of candidates. In focusing on modeling high-dimensional data, specifically those with a functional nature, this article proposes a method for selecting and estimating basis function coefficients based on the Bayesian paradigm. Compared to frequentist selection and regularization methods, which only provide point estimates, Bayesian techniques provide more information about the influence of candidate basis functions as they automatically return credibility intervals, enabling hypothesis tests.

Regarding the method proposed herein for selecting basis functions, it was shown that the model has enough layers to guarantee the definition of non-informative hyperparameters. Even though it presented results with low sensitivity to hyperparameters, the model can still guarantee the interpretability of the hyperparameter $\mu$ as the ratio between the expected prior number of bases to be selected and the total number of candidate bases.

We showed via simulation studies that our proposed methodology can correctly select the bases that should be included in the model. In addition, the proposed method can quantify the influence of each of the selected bases with satisfactory precision, consequently improving the fit of the response variable.

Even when compared to LASSO and Bayesian LASSO, the comparison had to be performed by evaluating a single curve. Unlike the proposed model, these methods were not built to contemplate the specificities inherent to the individual variability of each functional observation.

Compared to traditional curve-fitting models, the proposed model provides a more flexible description of the relationship between the functional curve and the basis functions, being extremely useful in configurations whose bases are orthogonal, such as Fourier bases and even functional principal components.

Although this article discusses curve modeling via the selection of basis functions, the methodology opens the way for developing similar Bayesian models focusing on the selection field of covariates in functional regression models.

%% file: appendix.tex
\appendix

	



 \section{Full conditional distributions}\label{A_full}

The model given in \eqref{modeloAuto} will be considered to calculate the full conditional distributions. The procedure is analogous to the model in \eqref{modeloRaiz}. It is noteworthy that whenever there is a vector whose notation specifies a negative subscript sign (e.g. $\vec{\beta_{.-i}}$ and $\vec{\theta_{-[ki]}}$), this vector must be interpreted as the version of the respective complete vector that does not contain the components associated with the negative index.

The full conditional distribution of the partial coefficients of each $i$-th curve, $\vec{\beta_{.i}}=(\beta_{1i},\beta_{2i},\dots,\beta_{Ki})^{'}$, is given by: 

\begin{gather}
	f(\vec{\beta_{.i}}|\vec{\beta_{.-i}},\vec{\theta},\vec{\mu},\sigma^2,\tau^2,\vec{Z},\vec{Y})\nonumber\\
	\propto\pi(\sigma^2)\pi(\tau^2)\pi(\vec{\mu})f(\vec{\theta}|\vec{\mu})f(\vec{\beta}|\sigma^2,\tau^2)p(\vec{Z}|\vec{\theta})f(\vec{Y}|\vec{\beta},\vec{Z},\sigma^2)\nonumber\\
	\propto f(\vec{\beta_{.i}}|\sigma^2,\tau^2)f(\vec{Y}|\vec{\beta},\vec{Z},\sigma^2)\nonumber\\
	\propto \left[\prod_{k=1}^{K}f(\beta_{ki}|\sigma^2,\tau^2)\right]\left[\prod_{s=1}^{m}\prod_{j=1}^{n_{s}}f(Y_{sj}|\vec{\beta},\vec{Z},\sigma^2)\right]\nonumber\\
	\propto \exp\left\{-\frac{\sum_{k=1}^{K}\beta_{ki}^{2}}{2\sigma^2\tau^2}\right\}
	\exp\left\{-\frac{\sum_{s=1}^{m}\sum_{j=1}^{n_{s}}(y_{sj}-g(t_{sj}))^2}{2\sigma^2}\right\}\nonumber\\
	\propto \exp\left\{-\frac{\sum_{k=1}^{K}\beta_{ki}^{2}}{2\sigma^2\tau^2}\right\}
	\exp\left\{-\frac{\sum_{j=1}^{n_{i}}(y_{ij}-g(t_{ij}))^2}{2\sigma^2}\right\}\nonumber\\
	\propto \exp\left\{-\frac{\vec{\beta_{.i}}^{'}\vec{\beta_{.i}}}{2\sigma^2\tau^2}-
	\frac{\vec{C_{i}}^{'}\vec{C_{i}}}{2\sigma^2}\right\}\text{,}\nonumber
\end{gather}where  $\vec{C_{i}}=(C_{i1},C_{i2},\dots,C_{in_{i}})$, with $C_{ij}=y_{ij} - g(t_{ij})$.
Recall that $g(t_{ij})=\sum_{k=1}^{K}\beta_{ki}Z_{ki}B_{k}(t_{ij})$, which is equivalent to $\vec{\beta_{.i}}^{'}\vec{G_{ij}}$, in which $\vec{G_{ij}}=(Z_{1i}B_{1}(t_{ij}),Z_{2i}B_{2}(t_{ij}),$ $\dots,Z_{Ki}B_{K}(t_{ij}))^{'}$. So, we have $C_{ij}=y_{ij}-\vec{\beta_{.i}}^{'}\vec{G_{ij}}$.
Thus, 
\begin{gather}
	f(\vec{\beta_{.i}}|\vec{\beta_{.-i}},\vec{\theta},\vec{\mu},\sigma^2,\tau^2,\vec{Z},\vec{Y})\nonumber\\
	\propto \exp\left\{-\frac{\frac{1}{\tau^2}\vec{\beta_{.i}}^{'}\vec{\beta_{.i}}+
		\vec{C_{i}}^{'}\vec{C_{i}}}{2\sigma^2}\right\}\nonumber\\
	\propto \exp\left\{-\frac{\frac{1}{\tau^2}\vec{\beta_{.i}}^{'}\vec{\beta_{.i}}+
		\sum_{j=1}^{n_{i}}(y_{ij}-\vec{\beta_{.i}}^{'}\vec{G_{ij}})^{'}(y_{ij}-\vec{\beta_{.i}}^{'}\vec{G_{ij}})}{2\sigma^2}\right\}\nonumber\\
	\propto \exp\left\{-\frac{\frac{1}{\tau^2}\vec{\beta_{.i}}^{'}\vec{\beta_{.i}}+
		\sum_{j=1}^{n_{i}}\left[y_{ij}^{2}-2y_{ij}\vec{\beta_{.i}}^{'}\vec{G_{ij}}+\vec{\beta_{.i}}^{'}\vec{G_{ij}}\vec{G_{ij}}^{'}\vec{\beta_{.i}}\right]}{2\sigma^2}\right\}\nonumber\\
	\propto \exp\left\{-\frac{\vec{\beta_{.i}}^{'}\left[\frac{1}{\tau^2}\vec{I}+\sum_{j=1}^{n_{i}}\vec{G_{ij}}\vec{G_{ij}}^{'}\right]\vec{\beta_{.i}}-2\vec{\beta_{.i}}^{'}\sum_{j=1}^{n_{i}}\vec{G_{ij}}y_{ij}}{2\sigma^2}\right\}\text{.}\nonumber
\end{gather}

\noindent By Cholesky decomposition, there is $\vec{D_i}=\left[\frac{1}{\tau^2}\vec{I}+\sum_{j=1}^{n_{i}}\vec{G_{ij}}\vec{G_{ij}}^{'}\right]=\vec{L_{i}}\vec{L_{i}}^{'}$, so that:
\begin{gather}
	f(\vec{\beta_{.i}}|\vec{\beta_{.-i}},\vec{\theta},\vec{\mu},\sigma^2,\tau^2,\vec{Z},\vec{Y})\nonumber\\
	\propto \exp\left\{-\frac{\vec{\beta_{.i}}^{'}\vec{L_{i}}\vec{L_{i}}^{'}\vec{\beta_{.i}}-2\vec{\beta_{.i}}^{'}\sum_{j=1}^{n_{i}}\vec{G_{ij}}y_{ij}}{2\sigma^2}\right\}\nonumber\\
	\propto \exp\left\{-\frac{(\vec{L_{i}}^{'}\vec{\beta_{.i}})^{'}\vec{L_{i}}^{'}\vec{\beta_{.i}}-2\vec{\beta_{.i}}^{'}\vec{L_{i}}\vec{L_{i}}^{-1}\sum_{j=1}^{n_{i}}\vec{G_{ij}}y_{ij}}{2\sigma^2}\right\}\nonumber\\
	\propto \exp\left\{-\frac{(\vec{L_{i}}^{'}\vec{\beta_{.i}})^{'}\vec{L_{i}}^{'}\vec{\beta_{.i}}-2(\vec{L_{i}}^{'}\vec{\beta_{.i}})^{'}\vec{L_{i}}^{-1}\sum_{j=1}^{n_{i}}\vec{G_{ij}}y_{ij}}{2\sigma^2}\right\}\nonumber\\
	\propto \exp\left\{-\frac{(\vec{L_{i}}^{'}\vec{\beta_{.i}}-\vec{L_{i}}^{-1}\sum_{j=1}^{n_{i}}\vec{G_{ij}}y_{ij})^{'}(\vec{L_{i}}^{'}\vec{\beta_{.i}}-\vec{L_{i}}^{-1}\sum_{j=1}^{n_{i}}\vec{G_{ij}}y_{ij})}{2\sigma^2}\right\}\nonumber\\
	\propto \exp\left\{-\frac{(\vec{L_{i}}^{'}\vec{\beta_{.i}}-\vec{L_{i}}^{-1}\sum_{j=1}^{n_{i}}\vec{G_{ij}}y_{ij})^{'}\vec{L_{i}}^{-1}\vec{L_{i}}\vec{L_{i}}^{'}(\vec{L_{i}}^{'})^{-1}(\vec{L_{i}}^{'}\vec{\beta_{.i}}-\vec{L_{i}}^{-1}\sum_{j=1}^{n_{i}}\vec{G_{ij}}y_{ij})}{2\sigma^2}\right\}\nonumber\\
	\propto \exp\left\{-\frac{(\vec{\beta_{.i}}-(\vec{L_{i}}\vec{L_{i}}^{'})^{-1}\sum_{j=1}^{n_{i}}\vec{G_{ij}}y_{ij})^{'}\vec{L_{i}}\vec{L_{i}}^{'}(\vec{\beta_{.i}}-(\vec{L_{i}}\vec{L_{i}}^{'})^{-1}\sum_{j=1}^{n_{i}}\vec{G_{ij}}y_{ij})}{2\sigma^2}\right\}\nonumber
\end{gather}
$\Longrightarrow$
\begin{gather}
f(\vec{\beta_{.i}}|\vec{\beta_{.-i}},\vec{\theta},\vec{\mu},\sigma^2,\tau^2,\vec{Z},\vec{Y})=f(\vec{\beta_{.i}}|{\sigma^2},{\tau^2},\vec{Z},\vec{Y})\nonumber\\
\propto \exp\left\{-\frac{(\vec{\beta_{.i}}-\vec{D_i}^{-1}\sum_{j=1}^{n_{i}}\vec{G_{ij}}y_{ij})^{'}(\vec{D_i}^{-1})^{-1}(\vec{\beta_{.i}}-\vec{D_i}^{-1}\sum_{j=1}^{n_{i}}\vec{G_{ij}}y_{ij})}{2\sigma^2}\right\}\text{.}
	\label{cmpos_beta}
\end{gather}

As a result, $f(\vec{\beta_{.i}}|\vec{\beta_{.-i}},\vec{\theta},\vec{\mu},\sigma^2,\tau^2,\vec{Z},\vec{Y})$ is precisely a multivariate normal distribution with mean equal to $\vec{D_i}^{-1}\sum_{j=1}^{n_{i}}\vec{G_{ij}}y_{ij}$ and variance $\sigma^2\vec{D_i}^{-1}$. The mean of this normal distribution can still be described in a simplified way when considering: 
\begin{equation}
	\vec{G_{i.}}=\begin{pmatrix}
		\vec{G_{i1}}\\
		\vec{G_{i2}}\\
		\vdots\\
		\vec{G_{in_{i}}}
	\end{pmatrix}=\begin{pmatrix}
		Z_{1i}B_{1}(t_{i1})&Z_{2i}B_{2}(t_{i1})&\dots&Z_{Ki}B_{K}(t_{i1})\\
		Z_{1i}B_{1}(t_{i2})&Z_{2i}B_{2}(t_{i2})&\dots&Z_{Ki}B_{K}(t_{i2})\\
		\vdots&\vdots&\ddots&\vdots\\
		Z_{1i}B_{1}(t_{in_{i}})&Z_{2i}B_{2}(t_{in_{i}})&\dots&Z_{Ki}B_{K}(t_{in_{i}})
	\end{pmatrix}\text{.}\nonumber
\end{equation}

\noindent Then we obtain: 
\begin{equation}
	\E(\vec{\beta_{.i}}|\vec{\beta_{.-i}},\vec{\theta},\vec{\mu},\sigma^2,\tau^2,\vec{Z},\vec{Y})=\vec{D_{i}}^{-1}\vec{G_{i.}}^{'}\vec{y_{i.}}\text{.}
	\label{eq:mean_beta_i}
\end{equation}

Similarly, we obtain the full conditional distributions of the $\theta_{ki}'s$, taking:

\begin{gather}
	f(\theta_{ki}|\vec{\beta},\vec{\theta_{-[ki]}},\vec{\mu},\sigma^2,\tau^2,\vec{Z},\vec{Y})\nonumber\\
	\propto 
	\pi(\sigma^2)\pi(\tau^2)\pi(\vec{\mu})f(\theta_{ki}|\mu_{ki})f(\vec{\theta_{-[ki]}}|\vec{\mu_{-[ki]}})f(\vec{\beta}|\sigma^2,\tau^2)p(\vec{Z}|\vec{\theta})f(\vec{Y}|\vec{\beta},\vec{Z},\sigma^2)\nonumber\\
	\propto f(\theta_{ki}|\mu_{ki}) f(Z_{ki}|\theta_{ki})\propto(\theta_{ki})^{\mu_{ki}-1}(1-\theta_{ki})^{(1-\mu_{ki})-1}(\theta_{ki})^{Z_{ki}}(1-\theta_{ki})^{1-Z_{ki}}\nonumber
\end{gather}
$\Longrightarrow$
\begin{gather}
f(\theta_{ki}|\vec{\beta},\vec{\theta_{-[ki]}},\vec{\mu},\sigma^2,\tau^2,\vec{Z},\vec{Y})=f(\theta_{ki}|\mu_{ki},Z_{ki})\nonumber\\
\propto(\theta_{ki})^{\mu_{ki}+Z_{ki}-1}(1-\theta_{ki})^{(1-\mu_{ki})-Z_{ki}+1-1}\text{.}
		\label{cmpos_theta}
\end{gather}

\noindent which proves that $f(\theta_{ki}|\vec{\beta},\vec{\theta_{-[ki]}},\vec{\mu},\sigma^2,\tau^2,\vec{Z},\vec{Y})$ is a beta density with the first parameter being $\mu_{ki}+Z_{ki}$ and the second being $2-Z_{ki}-\mu_{ki}$.

In order to obtain the full conditional distributions of the latent variables, it is necessary to calculate the probability: 
\begin{gather}
	\prob(Z_{ki}=1|\vec{\beta},\vec{\theta},\vec{\mu},\sigma^2,\tau^2,\vec{Z_{-[ki]}},\vec{Y})\nonumber\\[5pt]\scriptsize
	=\frac{\pi(\sigma^2)\pi(\tau^2)\pi(\vec{\mu})f(\vec{\theta}|\vec{\mu})f(\vec{\beta}|\sigma^2,\tau^2)\prob(Z_{ki}=1|\vec{\theta})\prob(\vec{Z_{-[ki]}}=z_{-[ki]}|\vec{\theta})f(\vec{Y}|\vec{\beta},Z_{ki}=1,\vec{Z_{-[ki]}}=z_{-[ki]},\sigma^2)}{\sum_{z=0}^{1}\pi(\sigma^2)\pi(\tau^2)\pi(\vec{\mu})f(\vec{\theta}|\vec{\mu})f(\vec{\beta}|\sigma^2,\tau^2)\prob(Z_{ki}=z|\vec{\theta})\prob(\vec{Z_{-[ki]}}=z_{-[ki]}|\vec{\theta})f(\vec{Y}|\vec{\beta},Z_{ki}=z,\vec{Z_{-[ki]}}=z_{-[ki]},\sigma^2)}\nonumber\\[5pt]\normalsize
	=\frac{\prob(Z_{ki}=1|\vec{\theta})\prob(\vec{Z_{-[ki]}}=z_{-[ki]}|\vec{\theta})f(\vec{Y}|\vec{\beta},Z_{ki}=1,\vec{Z_{-[ki]}}=z_{-[ki]},\sigma^2)}{\sum_{z=0}^{1}\prob(Z_{ki}=z|\vec{\theta})\prob(\vec{Z_{-[ki]}}=z_{-[ki]}|\vec{\theta})f(\vec{Y}|\vec{\beta},Z_{ki}=z,\vec{Z_{-[ki]}}=z_{-[ki]},\sigma^2)}\text{.}\nonumber
\end{gather} 

Simplifying this, we have: 
\begin{gather}
	=\frac{\theta_{ki}f(\vec{Y}|\vec{\beta},Z_{ki}=1,\vec{Z_{-[ki]}}=z_{-[ki]},\sigma^2)}{(1-\theta_{ki})f(\vec{Y}|\vec{\beta},Z_{ki}=0,\vec{Z_{-[ki]}}=z_{-[ki]},\sigma^2)+\theta_{ki}f(\vec{Y}|\vec{\beta},Z_{ki}=1,\vec{Z_{-[ki]}}=z_{-[ki]},\sigma^2)}\nonumber\\
	=\frac{\theta_{ki}}{(1-\theta_{ki})\frac{f(\vec{Y}|\vec{\beta},Z_{ki}=0,\vec{Z_{-[ki]}}=z_{-[ki]},\sigma^2)}{f(\vec{Y}|\vec{\beta},Z_{ki}=1,\vec{Z_{-[ki]}}=z_{-[ki]},\sigma^2)}+\theta_{ki}}\text{.}\nonumber
\end{gather}

\noindent So that $g_{0}(t_{ij})=\sum_{r\ne k}z_{ri}\beta_{ri}B_{r}(t_{ij})$ and $g_{1}(t_{ij})=\beta_{ki}B_{k}(t_{ij})+\sum_{r\ne k}z_{ri}\beta_{ri}B_{r}(t_{ij})$, then we obtain: 
\begin{gather}
	\frac{f(\vec{Y}|\vec{\beta},Z_{ki}=0,\vec{Z_{-[ki]}}=z_{-[ki]},\sigma^2)}{f(\vec{Y}|\vec{\beta},Z_{ki}=1,\vec{Z_{-[ki]}}=z_{-[ki]},\sigma^2)}\nonumber\\
	=\exp\Bigg\{\frac{1}{2\sigma^2}\Bigg[\sum_{j=1}^{n_{i}}(y_{ij}-g_{1}(t_{ij}))^2+\sum_{s\ne i}\sum_{j=1}^{n_{s}}(y_{sj}-g(t_{sj}))^2\nonumber\\-\sum_{j=1}^{n_{i}}(y_{ij}-g_{0}(t_{ij}))^2-\sum_{s\ne i}\sum_{j=1}^{n_{s}}(y_{sj}-g(t_{sj}))^2\Bigg]\Bigg\}\nonumber\\
	=\exp\Bigg\{\frac{1}{2\sigma^2}\Bigg[\sum_{j=1}^{n_{i}}(y_{ij}-g_{1}(t_{ij}))^2-\sum_{j=1}^{n_{i}}(y_{ij}-g_{0}(t_{ij}))^2\Bigg]\Bigg\}\text{.}\nonumber
\end{gather}

\noindent Thus,
\begin{gather}
	\prob(Z_{ki}=1|\vec{\beta},\vec{\theta},\vec{\mu},\sigma^2,\tau^2,\vec{Z_{-[ki]}},\vec{Y})=\prob(Z_{ki}=1|\vec{\beta},\theta_{ki},{\sigma^2},\vec{Z_{-[ki]}},\vec{Y})\nonumber\\
	=\frac{\theta_{ki}}{(1-\theta_{ki})\exp\Bigg\{\frac{1}{2\sigma^2}\Bigg[\sum_{j=1}^{n_{i}}(y_{ij}-g_{1}(t_{ij}))^2-\sum_{j=1}^{n_{i}}(y_{ij}-g_{0}(t_{ij}))^2\Bigg]\Bigg\}+\theta_{ki}}\text{.}
	\label{cmpos_Z}
\end{gather}

The construction of the full conditional distribution of $\tau^{2}$ is straightforward. Taking into account that: 
\begin{gather}
	f(\tau^2|\vec{\beta},\vec{\theta},\vec{\mu},\sigma^2,\vec{Z},\vec{Y})\propto 
	\pi(\sigma^2)\pi(\tau^2)\pi(\vec{\mu})f(\vec{\theta}|\vec{\mu})f(\vec{\beta}|\sigma^2,\tau^2)p(\vec{Z}|\vec{\theta})f(\vec{Y}|\vec{\beta},\vec{Z},\sigma^2)\nonumber\\
	\propto\pi(\tau^2)\prod_{i=1}^{m}\prod_{k=1}^{K}f(\beta_{ki}|\sigma^2,\tau^2)\propto\left(\frac{1}{\tau^2}\right)^{\lambda_{1}+1}\exp\left\{-\frac{\lambda_{2}}{\tau^2}\right\}\prod_{i=1}^{m}\prod_{k=1}^{K}\left(\frac{1}{\sigma^2\tau^2}\right)^{\frac{1}{2}}\exp\left\{-\frac{\beta_{ki}^2}{2\sigma^2\tau^2}\right\}\nonumber
\end{gather}
$\Longrightarrow$
\begin{gather}
	f(\tau^2|\vec{\beta},\vec{\theta},\vec{\mu},\sigma^2,\vec{Z},\vec{Y})=f(\tau^2|\vec{\beta},{\sigma^2})\nonumber\\
	\propto\left(\frac{1}{\tau^2}\right)^{\frac{mK}{2}+\lambda_{1}+1}\exp\left\{-\frac{\sum_{i=1}^{m}\sum_{k=1}^{K}\frac{\beta_{ki}^{2}}{\sigma^2}+2\lambda_{2}}{2\tau^2}\right\}\text{,}
	\label{cmpos_tau2}
\end{gather}it can be seen that the distribution of $\tau^{2}$ conditioned to all the rest is an inverse gamma distribution with the first parameter being determined by $\frac{mK}{2}+\lambda_{1}$ and the second by $\frac{\sum_{i=1}^{m}\sum_{k=1}^{K}\frac{\beta_{ki}^{2}}{\sigma^2}+2\lambda_{2}}{2}$.

Finally, the other variance component of the model is left to be calculated. It is known that:
\begin{gather}
	f(\sigma^2|\vec{\beta},\vec{\theta},\vec{\mu},\tau^2,\vec{Z},\vec{Y})\propto 
	\pi(\sigma^2)\pi(\tau^2)\pi(\vec{\mu})f(\vec{\theta}|\vec{\mu})f(\vec{\beta}|\sigma^2,\tau^2)p(\vec{Z}|\vec{\theta})f(\vec{Y}|\vec{\beta},\vec{Z},\sigma^2)\nonumber\\
	\propto\pi(\sigma^2)f(\vec{\beta}|\sigma^2,\tau^2)f(\vec{Y}|\vec{\beta},\vec{Z},\sigma^2)\propto\pi(\sigma^2)\left[\prod_{i=1}^{m}\prod_{k=1}^{K}f(\beta_{ki}|\sigma^2,\tau^2)\right]\left[\prod_{i=1}^{m}\prod_{j=1}^{n_{i}}f(Y_{ij}|\vec{\beta},\vec{Z},\sigma^2)\right]\nonumber\\
	\propto\left(\frac{1}{\sigma^2}\right)^{\delta_{1}+1}\exp\left\{-\frac{\delta_{2}}{\sigma^2}\right\}\left[\prod_{i=1}^{m}\prod_{k=1}^{K}\left(\frac{1}{\sigma^2\tau^2}\right)^{\frac{1}{2}}\exp\left\{-\frac{\beta_{ki}^2}{2\sigma^2\tau^2}\right\}\right]\nonumber\\\times\left[\prod_{i=1}^{m}\prod_{j=1}^{n_{i}}\left(\frac{1}{\sigma^2}\right)^{\frac{1}{2}}\exp\left\{-\frac{(y_{ij}-g(t_{ij}))^2}{2\sigma^2}\right\}\right]\nonumber
\end{gather}
$\Longrightarrow$
\begin{gather}
	f(\sigma^2|\vec{\beta},\vec{\theta},\vec{\mu},\tau^2,\vec{Z},\vec{Y})=f(\sigma^2|\vec{\beta},{\tau^2},\vec{Z},\vec{Y})\nonumber\\
\propto\left(\frac{1}{\sigma^2}\right)^{\frac{\sum_{i=1}^{m}n_{i}}{2}+\frac{mK}{2}+\delta_{1}+1}\exp\left\{-\frac{\sum_{i=1}^{m}\sum_{j=1}^{n_{i}}(y_{ij}-g(t_{ij}))^{2}+\sum_{i=1}^{m}\sum_{k=1}^{K}\frac{\beta_{ki}^2}{\tau^2}+2\delta_{2}}{2\sigma^2}\right\}\text{.}
	\label{cmpos_sigma2}
\end{gather}

Thus, it is concluded that, as for $\tau^{2}$, the full conditional distribution of $\sigma^{2}$ is an inverse gamma, but with the shape parameter being determined by $\frac{\sum_{i=1}^{m}n_{i}}{2}+\frac{mK}{2}+\delta_{1}$ and the rate by $\frac{\sum_{i=1}^{m}\sum_{j=1}^{n_{i}}(y_{ij}-g(t_{ij}))^{2}+\sum_{i=1}^{m}\sum_{k=1}^{K}\frac{\beta_{ki}^2}{\tau^2}+2\delta_{2}}{2}$.
An additional account must be performed when considering $\mu_{ki}$ as parameter: 

\begin{gather}
	f(\mu_{ki}|\vec{\beta},\vec{\theta},\vec{\mu_{-[ki]}},\sigma^2,\tau^2,\vec{Z},\vec{Y})\nonumber\\
	\propto 
	\pi(\sigma^2)\pi(\tau^2)\pi(\mu_{ki})\pi(\vec{\mu_{-[ki]}})f(\theta_{ki}|\mu_{ki})f(\vec{\theta_{-[ki]}}|\vec{\mu_{-[ki]}})f(\vec{\beta}|\sigma^2,\tau^2)p(\vec{Z}|\vec{\theta})f(\vec{Y}|\vec{\beta},\vec{Z},\sigma^2)\nonumber
\end{gather}
$\Longrightarrow$
\begin{gather}
f(\mu_{ki}|\vec{\beta},\vec{\theta},\vec{\mu_{-[ki]}},\sigma^2,\tau^2,\vec{Z},\vec{Y})=f(\mu_{ki}|\theta_{ki})\nonumber\\
\propto \pi(\mu_{ki})f(\theta_{ki}|\mu_{ki})\propto I_{(0,\psi)}(\mu_{ki})\theta_{ki}^{\mu_{ki}}(1-\theta_{ki})^{1-\mu_{ki}}\text{.}
	\label{cmpos_mu}
\end{gather}Thus, $f(\mu_{ki}|\vec{\beta},\vec{\theta},\vec{\mu_{-[ki]}},\sigma^2,\tau^2,\vec{Z},\vec{Y})$ is a continuous bernoulli distribution with parameter $\theta_{ki}$ truncated at range $(0,\psi)$.